\documentclass[11pt,a4paper]{article}
\usepackage[utf8]{inputenc}
\usepackage{amsmath, amssymb, bm}
\usepackage{geometry}
\usepackage{hyperref}
\usepackage{xcolor}
\usepackage{authblk}
\usepackage{float}
\usepackage{graphicx}

\hypersetup{
    colorlinks=true,
    linkcolor=blue!70!black,
    urlcolor=blue!70!black
}

\title{Origin of effective non-Fourier heat conduction phenomena in heterogeneous materials}
\author[1,2]{Róbert Kovács}
\date{\today}

\affil[1]{Department of Energy Engineering, Faculty of Mechanical Engineering, Budapest University of Technology and Economics, Műegyetem rkp. 3., H-1111 Budapest, Hungary}
\affil[2]{Department of Theoretical Physics, Wigner Research Centre for Physics, Institute for Particle and Nuclear Physics, Budapest, Hungary}

\begin{document}
\maketitle

\begin{abstract}
Phenomenological models of non-Fourier heat conduction often lack a strict microstructural foundation, leading to ambiguities when modeling complex heterogeneous materials. In this study, we derive a continuum heat equation beyond Fourier's law using spatial volume averaging for a two-component system. We analytically prove that the experimentally observed static and dynamic thermal diffusivity arise directly from the distinct material properties, concluding that heterogeneous media are inherently over-diffusive. The resulting heat equation is thermodynamically compatible, and the microstructural origin allows the calculation of non-Fourier transport coefficients. Furthermore, we demonstrate that finite-sample boundaries introduce higher-order spatial non-localities, thereby explaining the size dependence of over-diffusion. We validate the model against experimental data across metal and carbon foams, rocks, and metal-organic frameworks.
\end{abstract}

\section{Introduction}

Although Fourier's law of heat conduction is one of the most well-known and successful models in continuum physics, its limitations become pronounced in advanced engineering materials \cite{YangEtal21}. While non-Fourier phenomena, such as phonon hydrodynamics and second sound, were historically observed only in high-purity crystals at cryogenic temperatures \cite{MulRug98, JacWal71}, recent experimental advances have revealed a so-called over-diffusive deviation at room temperature in macroscopically sized, highly heterogeneous materials \cite{FehEtal21, FehEtal24}. Materials with complex internal structures, such as metal foams, open-cell carbon foams, composites, and porous rocks, exhibit notable deviations from classical diffusion during transient processes \cite{FehEtal24b}. Although heat transport may locally obey Fourier's law at the microscopic pore scale, the distinct thermophysical properties of the constituents, complex interfacial phenomena, and the complex geometrical arrangement can induce macroscopically observable temporal memory and spatial non-locality.

In order to rigorously describe these macroscopic effects without relying on the phenomenological use of internal variables \cite{BerJurMau11, VanEtal08}, we aim to derive an effective evolution equation that models the macroscopic behavior of heterogeneous materials and explains the experimental data gathered.
The volume averaging method can be widely utilized for heterogeneous materials to determine their effective (or apparent) transport properties \cite{QuinWhit88, QuinWhit93, QuinWhit94}. This is a mathematical upscaling technique that transitions microscopic, pore-level differential equations into a continuous macroscopic representation based on the given microscopic structure. The application of this method fundamentally relies on the separation of length scales, i.e., distinguishing the microscopic characteristic length from the macroscopic scale of the material structure. The upscaling procedure utilizes spatial smoothing theorems, which separate a local field into an intrinsic macroscopic average and a microscopic spatial fluctuation ($T = \langle T\rangle + \tilde{T}$). Overall, a closure problem is formulated to map the microscopic fluctuations to macroscopic driving forces via periodic boundary value problems, enabling the deterministic calculation of effective transport tensors and volumetric heat transfer coefficients.

When heterogeneous media are subjected to rapid thermal transients, the assumption of local thermal equilibrium may no longer hold, especially for complexly structured materials. While numerous non-equilibrium irreversible thermodynamic frameworks have been developed in recent decades -- such as extended irreversible thermodynamics \cite{CimmEtal14, Lebon2017}, rational extended thermodynamics \cite{MulRug98}, GENERIC \cite{PavEtal18b, Grmela2018b}, and internal variables \cite{ColGur67, BerVan17b} -- the local thermal non-equilibrium (LTNE) framework \cite{DuvalEtal04, PatiEtal22, GandoGray18, Sobolev97} can also be a valid approach to derive heat equations beyond Fourier. The LTNE approach yields a 2-temperature (2T) model in which the balance equations of internal energy  for the individual components are coupled \cite{Sobolev97, Sobolev16}; for instance, the one-dimensional (macroscopic) internal energy balance for phase $\alpha$ is given by
\begin{align}
\varepsilon_\alpha (\rho c)_\alpha \frac{\partial T_\alpha}{\partial t} = k_\alpha \frac{\partial^2 T_\alpha}{\partial x^2} - H (T_\alpha - T_\beta),
\end{align}
where $T_\alpha$ and $T_\beta$ are the intrinsic phase temperatures, $k_\alpha$ is the thermal conductivity, and $H$ is the interfacial heat transfer coefficient. In regard to the specific heat $c$, we note that $c_v=c_p=c$ as thermal expansion is neglected, hence we assume that there is no difference between the isochoric and isobaric specific heat for each phase. Additionally, $\rho$ is the mass density, and $\partial_t$ (or $\partial_x$) denotes the corresponding partial derivative with respect to time ($t$) or space ($x$). Furthermore, $\varepsilon_\alpha$ denotes the volume ratio of phase $\alpha$. The coupled internal energy balance for phase $\beta$ follows symmetrically as 
\begin{align}
    \varepsilon_\beta (\rho c)_\beta \frac{\partial T_\beta}{\partial t} = k_{\beta} \frac{\partial^2 T_\beta}{\partial x^2} + H (T_\alpha - T_\beta).
\end{align}
Although such a 2T approach can capture complex behavior, and its $T$-representation (i.e., rearranging the entire system of evolution equations to one of the temperatures) resembles non-Fourier equations, in a one-dimensional form,
\begin{align}
 \tau \frac{\partial^2 T_\alpha}{\partial t^2} + \frac{\partial T_\alpha}{\partial t} = \Lambda \frac{\partial^2 T_\alpha}{\partial x^2}  + l^2 \frac{\partial^3 T_\alpha}{\partial t\partial x^2} - \xi \frac{\partial^4 T_\alpha}{\partial x^4},
\end{align}
with strictly positive macroscopic transport parameters of $\tau$, $\Lambda$, $l$, and $\xi$, the 2T model is strictly specified for heterogeneous materials as it can model diffusion only despite exhibiting wave-like relaxation behavior mathematically. Its parameters are defined by
\begin{align}
   \tau = \frac{C_\alpha C_\beta}{H(C_\alpha + C_\beta)}, \quad \Lambda = \frac{C_\alpha a_\alpha + C_\beta a_\beta}{C_\alpha + C_\beta}, \quad l^2 = \tau (a_\alpha + a_\beta), \quad \xi = \tau a_\alpha a_\beta, 
\end{align}
in which $C_i =\varepsilon_i\rho_i c_i$, and $a$ denotes the thermal diffusivity, $a_i= k_i/C_i$ for $i=\{\alpha, \beta\}$. In the present paper, we combine the volume averaging method with the 2T approach in order to achieve a particular type of coupling between the components of the heterogeneous solid. However, when formulating such generalized heat equations, ensuring their thermodynamic compatibility is essential. Phenomenological approaches, such as Tzou's dual-phase-lag model \cite{Tzou95}, which relies on arbitrary Taylor-series expansions of the phase-lagged relation $\mathbf{q}(\mathbf{x}, t+\tau_q) = -k \nabla T(\mathbf{x}, t+\tau_T^{})$, has been proven to violate the Clausius-Duhem inequality \cite{Fabetal14, FabLaz14a, Quin07, ChiritaEtal17}, although it can formally reconstruct the $T$-representation of various heat equations beyond Fourier \cite{Kov24}. 
However, the constructive use of the second law of thermodynamics can guarantee stability, and thus the existence of a stable equilibrium \cite{VanFul12}.

This manuscript details the derivation of a macroscopic heat conduction equation, which is -- surprisingly -- expressed as a combination of Jeffreys and non-local heat equations, partially resembling the well-known Guyer--Krumhansl equation \cite{GK66b}. However, it is crucial to emphasize that any resemblance cannot be directly translated to either phonon hydrodynamics of rational extended thermodynamics or kinetic theory approaches, because the present framework exclusively uses continuum thermodynamics. The primary novelty of this work lies in demonstrating how the microscale coupling of different heat transfer channels obeying Fourier's law can macroscopically lead to an effective non-Fourier heat equation. The paper begins with the upscaling of the continuum, deriving the effective macroscopic transport equation for general two-component heterogeneous media. It is followed by an investigation of thermodynamic compatibility using the Clausius-Duhem inequality to reveal constraints that ensure positive entropy production. Finally, we study the reduction of this generalized system to highlight the role of static and dynamic timescales, defining the theoretical basis of the over-diffusive regime in which the dynamic timescale is larger than the static (Fourier) timescale. We compare our findings to recent experiments performed on various foams, rocks, and metal-organic frameworks. Additionally, we provide a brief overview of biological applications, showing how blood perfusion can modify the dynamics of heterogeneous material structures.

\section{The role of the representative volume element}

In order to systematically upscale the microscopic behavior of a heterogeneous material, we impose only minimal assumptions. We consider a heterogeneous two-phase material consisting of rigid, stationary domains, denoted as phase $\alpha$ and phase $\beta$, corresponding to a solid matrix and a saturating fluid, respectively. Furthermore, the macroscopic medium is isotropic, with a periodically repeating microstructure and uniformly distributed pores.

The foundation of this continuum approach relies on the separation of length scales following Gray \cite{Gray1975}, which is mathematically expressed through the inequality
\begin{align}
l \ll L \ll L_{\textrm{mac}}.
\end{align}
In this spatial hierarchy, $l$ denotes the characteristic microscopic length scale, such as the average pore diameter or grain size. The parameter $L$ represents the characteristic length scale of the representative volume element (RVE), which defines a theoretical averaging window that is sufficiently large to capture a statistically representative sample of the complex microstructure, yet small enough to be treated as a localized differential point within the macroscopic continuum, usually interpreted as a mesoscale. Finally, $L_{\textrm{mac}}$ corresponds to the macroscopic length scale. This scale separation and averaging ensure that an effective, homogenized continuum description exists independently of the specific complex structural topology of individual microscopic pores.

\begin{figure}[H]
    \centering
    \includegraphics[width=0.7\linewidth]{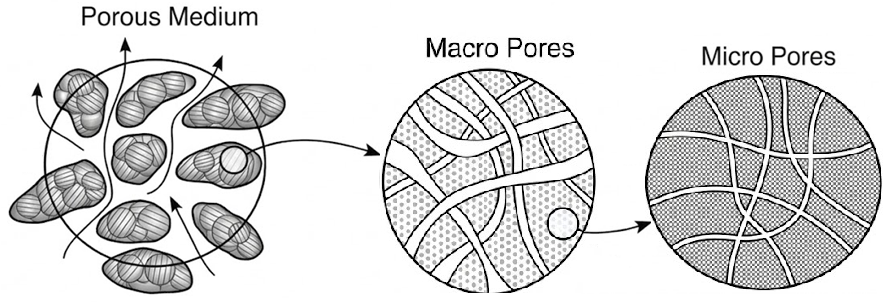}
    \caption{The schematics of various length scales in a heterogeneous material, based on \cite{QuinWhit93}.}
    \label{fig:1}
\end{figure}

\subsection{Microscopic governing equations}

At the microscopic scale within the pore domain, we assume classical heat conduction driven by Fourier's law. For phases $\alpha$ and $\beta$, the local balance of internal energy has the form
\begin{align}
(\rho c)_\alpha \frac{\partial T_{\alpha \rm m}}{\partial t} &= - \nabla \cdot \mathbf q_{\alpha \rm  m} = \nabla \cdot (k_{\alpha \rm  m} \nabla T_{\alpha \rm m}) = k_{\alpha \rm  m}\Delta T_{\alpha \rm  m}, \\
(\rho c)_\beta \frac{\partial T_{\beta \rm  m}}{\partial t} &= -\nabla \cdot \mathbf{q}_{\beta \rm  m} =\nabla \cdot (k_{\beta \rm m} \nabla T_{\beta \rm  m}) = k_{\beta \rm  m} \Delta T_{\beta \rm  m},
\end{align}
in which $\Delta$ denotes the Laplacian, $\nabla \cdot$ is the divergence, and $\nabla$ stands for the gradient. Furthermore, the subscript $\rm m$ denotes a microscopic variable to distinguish it from its upscaled continuum (macroscopic) counterpart. The thermal conductivity and the volumetric heat capacity of the respective phase are treated as constants. At the microscopic interface $A_{\alpha\beta}$ separating the two phases, the standard continuity conditions for temperature and normal heat flux are applied to ensure energy conservation.

\subsection{The volume averaging framework}

We employ spatial volume averaging over the representative volume element (RVE). We define two primary averaging operators for any given microscopic variable $\psi_{\alpha \rm m}$ defined in phase $\alpha$. The first one is evaluated over the total volume of the RVE, $V$, where $V_\alpha$ represents the volume occupied by phase $\alpha$, is defined as
\begin{align}
\langle \psi_{\alpha \rm m} \rangle = \frac{1}{V} \int_{V_\alpha} \psi_{\alpha \rm m} \, \textrm{d}V,
\end{align}
whereas the intrinsic volume average, evaluated solely over the spatial domain actually occupied by phase $\alpha$, is given by
\begin{align}
\langle \psi_{\alpha \rm m} \rangle^\alpha = \frac{1}{V_\alpha} \int_{V_\alpha} \psi_{\alpha \rm m} \, \textrm{d}V.
\end{align}
These two mathematical representations are linked by the phase volume fraction, defined as $\varepsilon_\alpha = V_\alpha / V$, which yields the identity
\begin{align}
\langle \psi_{\alpha \rm m} \rangle = \varepsilon_\alpha \langle \psi_{\alpha \rm m} \rangle^\alpha.
\end{align}

For the upscaling, it is necessary to commute the volume integration with spatial differential operators. This is achieved via the spatial averaging theorem \cite{Neum77}, which, for the spatial gradient of a variable, leads to
\begin{align}
\langle \nabla \psi_{\alpha \rm m} \rangle = \nabla \langle \psi_{\alpha \rm m} \rangle + \frac{1}{V} \int_{A_{\alpha\beta}} \mathbf{n}_{\alpha\beta} \psi_{\alpha \rm m} \, \textrm{d}A,
\end{align}
where $\mathbf{n}_{\alpha\beta}$ is the outward-pointing unit normal vector oriented from phase $\alpha$ into phase $\beta$. The resulting surface integral represents the energy exchange occurring directly at the phase interface.
In order to resolve the integration of microscopic variables across the boundary of the RVE, we introduce Gray's spatial decomposition \cite{Gray1975}, which splits the microscopic variable into a macroscopic average $T_\alpha \equiv \langle T_{\alpha \rm m} \rangle^\alpha$ and a local, microscopic spatial deviation $\tilde{T}_{\alpha \rm m}$ via the relation
\begin{align}
T_{\alpha \rm m} = T_\alpha + \tilde{T}_{\alpha \rm m}.
\end{align}
After substituting this decomposition into the surface integral of the spatial averaging theorem, the macroscopic intrinsic temperature $T_\alpha$ acts as a constant with respect to the local, sub-RVE pore-scale integration by definition. Therefore, it can be taken out of the interfacial surface integral, leading to
\begin{align}
\frac{1}{V} \int_{A_{\alpha\beta}} \mathbf{n}_{\alpha\beta} T_{\alpha \rm m} \, \textrm{d}A = T_\alpha \left[ \frac{1}{V} \int_{A_{\alpha\beta}} \mathbf{n}_{\alpha\beta} \, \textrm{d}A \right] + \frac{1}{V} \int_{A_{\alpha\beta}} \mathbf{n}_{\alpha\beta} \tilde{T}_{\alpha \rm m} \, \textrm{d}A.
\end{align}
According to the spatial averaging theorem, the interfacial surface integral of the normal vector over the closed phase interface is directly linked to the spatial gradient of the volume fraction according to
\begin{align}
\frac{1}{V} \int_{A_{\alpha\beta}} \mathbf{n}_{\alpha\beta} \, \textrm{d}A = -\nabla \varepsilon_\alpha.
\end{align}
Consequently, the term containing the macroscopic temperature $T_\alpha$ vanishes identically only under the condition that $-\nabla \varepsilon_\alpha = \mathbf{0}$; in other words, we restrict ourselves to heterogeneous materials that possess a spatially uniform porosity at the macroscopic scale. Under this assumption, the scaling between the microscopic and the macroscopic parts is solely determined by the spatial deviation fields $\tilde{T}_{\alpha \rm m}$.

\section{Transition from microscale to macroscale}

\subsection{Macroscopic energy balance for the mixture temperature}

It is necessary to formally derive the effective energy balance equation for the mixture temperature $T$. Let us recall the balances of the internal energy component-wise, 
\begin{align}
    C_\alpha \frac{\partial T_\alpha}{\partial t} &= -\nabla \cdot \mathbf{q}_\alpha - q_{\textrm{int}}, \label{eq:51}\\
    C_\beta \frac{\partial T_\beta}{\partial t} &= -\nabla \cdot \mathbf{q}_\beta + q_{\textrm{int}}, \label{eq:52}
\end{align}
and add Eq.~\eqref{eq:52} to Eq.~\eqref{eq:51},
\begin{equation}
    C_\alpha \frac{\partial T_\alpha}{\partial t} + C_\beta \frac{\partial T_\beta}{\partial t} = -\nabla \cdot (\mathbf{q}_\alpha + \mathbf{q}_\beta). \label{eq:53}
\end{equation}
We now recall the definition of the heat capacity-weighted macroscopic mixture temperature $T = (C_\alpha T_\alpha + C_\beta T_\beta)/C_{\textrm{eff}}$, and recognizing that the total macroscopic heat flux is the superposition of the phase fluxes, Eq.~\eqref{eq:53} reduces to
\begin{equation}
    C_{\textrm{eff}} \frac{\partial T}{\partial t} = -\nabla \cdot \mathbf{q}. \label{eq:macro_energy_balance}
\end{equation}
This derivation confirms that the standard form of the continuum energy balance is preserved for the macroscopic mixture temperature $T$.

In classical Local Thermal Non-Equilibrium (LTNE) models, the interfacial heat exchange is traditionally assumed to be driven solely by the local temperature difference at the macroscopic point $\mathbf{x}$, yielding 
\begin{align}
    q_{\textrm{int}}(\mathbf{x}, t) = H \left( T_\alpha(\mathbf{x},t) - T_\beta(\mathbf{x},t) \right).
\end{align}
However, this local formulation relies on the separation of length scales ($L \ll L_{\text{mac}}$). In highly heterogeneous materials (e.g., syntactic metal foams \cite{MarOrb23, KarEtal22}), the required RVE size $L$ can be large, often approaching the continuum scale $L_{\text{mac}}$. Because the thermal exchange occurs over a spatially distributed interfacial area $A_{\alpha\beta}$, the lack of scale separation means that the temperature gradients ($\nabla T_\alpha, \nabla T_\beta$) can vary significantly across the RVE. Consequently, the local heat exchange might need a non-local extension.

To rigorously capture the finite-size boundary constraints, we must account for the spatial distribution of the temperature field within the RVE. Assuming that the macroscopic temperature fields are sufficiently smooth over the RVE, expanding the local intrinsic temperature difference, $\delta T(\mathbf{x} + \mathbf{y},t) = T_\alpha(\mathbf{x} + \mathbf{y},t) - T_\beta(\mathbf{x} + \mathbf{y},t)$ with $\mathbf{y}$ denoting the local microscale position vector, using a Taylor series approximation up to the second-order yields
\begin{align}
\delta T(\mathbf{x} + \mathbf{y}, t) \approx \delta T(\mathbf{x},t) + \mathbf{y} \cdot \nabla [\delta T(\mathbf{x},t)] + \frac{1}{2} (\mathbf{y} \otimes \mathbf{y}) : \nabla \otimes\nabla [\delta T(\mathbf{x},t)]. \label{eq:taylor_expansion}
\end{align}
The total heat exchange within the RVE is proportional to the integral of this distributed temperature difference over the interface $A_{\alpha\beta}$. When we perform the spatial averaging of Equation \eqref{eq:taylor_expansion} over a statistically isotropic structure, the linear asymmetric term integrates to zero ($\langle \mathbf{y} \rangle_{A_{\alpha\beta}} = \mathbf{0}$). 
Conversely, the second-order term $\langle \mathbf{y} \otimes \mathbf{y} \rangle_{A_{\alpha\beta}}$ does not vanish; for an isotropic medium, the dyadic product $\mathbf{y} \otimes \mathbf{y}$ evaluates to a strictly positive scalar constant multiplied by the identity tensor, expressed as $l_{\textrm{nl}}^2 \mathbf{I}$; hence $\mathbf{I} : \nabla \otimes\nabla = \text{Tr}(\nabla \otimes \nabla) = \Delta$. Here, $l_{\textrm{nl}}$ is a non-local length scale, which can depend on the characteristic pore size.
Consequently, the integration yields a non-local interfacial heat exchange term defined by
\begin{align}
q_{\textrm{int}} &= H \left[ (T_\alpha - T_\beta) + l_{\textrm{nl}}^2 \Delta (T_\alpha - T_\beta) \right]. \label{eq:nonlocal_interact}
\end{align}

Let us now recall the evolution equations for the 
macroscopic intrinsic phase temperatures, also including the interaction between the two phases,
\begin{align}
C_\alpha \frac{\partial T_\alpha}{\partial t} &= - \nabla \cdot \mathbf q_{\alpha} - q_{\textrm{int}}, \label{eq:phase_a} \\
C_\beta \frac{\partial T_\beta}{\partial t} &= - \nabla \cdot \mathbf q_{\beta} + q_{\textrm{int}}. \label{eq:phase_b}
\end{align}

At the microscopic level, the heat fluxes within the individual phases are governed by classical Fourier's law, expressed as $\mathbf{q}_{\alpha \rm m} = -k_{\alpha \rm m} \nabla T_{\alpha \rm m}$ and $\mathbf{q}_{\beta \rm m} = -k_{\beta \rm m} \nabla T_{\beta \rm m}$. The total effective macroscopic heat flux $\mathbf{q}$ passing through the representative volume element is defined as the superposition of the volume averages of these microscopic point fluxes, expressed as
\begin{align}
\mathbf{q} &= \langle \mathbf{q}_{\alpha \rm m} \rangle + \langle \mathbf{q}_{\beta \rm m} \rangle. \label{eq:q_superficial_sum}
\end{align}
Focusing first on the phase $\alpha$ and assuming a constant microscopic thermal conductivity $k_{\alpha \rm m}$, the volume average evaluates to 
\begin{align}
    \langle \mathbf{q}_{\alpha \rm m} \rangle = -k_\alpha \langle \nabla T_{\alpha \rm m} \rangle
\end{align}
In order to evaluate this averaged spatial gradient, we must apply the spatial averaging theorem to commute the volume integration and the gradient operator, leading to
\begin{align}
\langle \nabla T_{\alpha \rm m} \rangle &= \varepsilon_\alpha \nabla T_\alpha + \frac{1}{V} \int_{A_{\alpha\beta}} \mathbf{n}_{\alpha\beta} T_{\alpha \rm m} \, \textrm{d}A,
\end{align}
where $T_\alpha = \langle T_{\alpha \rm m} \rangle^\alpha$ is the macroscopic intrinsic volume average of the phase temperature. Utilizing Gray's spatial decomposition again ($T_{\alpha \rm m} = T_\alpha + \tilde{T}_{\alpha \rm m}$), we substitute it directly into the surface integral. Since the macroscopic field $T_\alpha$ acts as a constant with respect to the microscopic integration, the term containing $T_\alpha$ vanishes due to the spatially uniform porosity expressed earlier. It simplifies the averaged gradient to
\begin{align}
\langle \nabla T_{\alpha \rm m} \rangle &= \varepsilon_\alpha \nabla T_\alpha + \frac{1}{V} \int_{A_{\alpha\beta}} \mathbf{n}_{\alpha\beta} \tilde{T}_{\alpha \rm m} \, \textrm{d}A. \label{eq:grad_T_alpha}
\end{align}
In order to interpret and evaluate the remaining surface integral in Equation \eqref{eq:grad_T_alpha}, we need to apply the so-called geometrical vector field $\mathbf{b}$, describing the shape of the microscale heterogeneity, and how its existence in the material distorts the homogeneous temperature field, following \cite{QuinWhit93}, and leading to $\tilde{T}_{\alpha \rm m} = \mathbf{b}_\alpha \cdot \nabla T_\alpha$. In other words, the vector field $\mathbf{b}$ maps the local microscopic temperature variation to the macroscopic temperature gradient, thereby achieving upscaling. Substituting this geometric closure relation back into the surface integral yields the closed-form macroscopic flux,
\begin{align}
\langle \mathbf{q}_{\alpha \rm m} \rangle &= - \underbrace{\left[ k_{\alpha \rm m} \left( \varepsilon_\alpha \mathbf{I} + \frac{1}{V} \int_{A_{\alpha\beta}} \mathbf{n}_{\alpha\beta} \mathbf{b}_\alpha \, \textrm{d}A \right) \right]}_{\mathbf{k}_{\alpha}} \cdot \nabla T_\alpha = - \mathbf{k}_{\alpha} \cdot \nabla T_\alpha.
\end{align}
The tensorial quantity within the square brackets represents the geometrically effective macroscopic conductivity tensor, $\mathbf{k}_{\alpha}$. However, assuming a statistically isotropic microstructure where the geometric effects vanish such that the tensor $\mathbf{k}_{\alpha}$ simplifies to a scalar $k_\alpha$, the macroscopic flux carried by the phase reduces to $\langle \mathbf{q}_{\alpha \rm m} \rangle = -k_\alpha \nabla T_\alpha$. Applying an identical spatial averaging and structural closure procedure to the phase $\beta$ yields the corresponding macroscopic flux, $\langle \mathbf{q}_{\beta \rm m} \rangle = -k_\beta \nabla T_\beta$. Superposing these phases, the total macroscopic heat flux reads as
\begin{align}
\mathbf{q} &= -\Big(k_\alpha \nabla T_\alpha + k_\beta \nabla T_\beta \Big ). \label{eq:def_q}
\end{align}

\subsection{Deriving the effective macroscopic heat equation}

In order to obtain a macroscopically effective model without containing any of the phase temperatures and separate phase equations, we need to introduce the macroscopic temperature $T$,
\begin{align}
    T = \frac{C_\alpha T_\alpha + C_\beta T_\beta}{C_{\textrm{eff}}}, \quad  C_{\textrm{eff}} = C_\alpha + C_\beta, \label{eq:def_T}
\end{align}
in which $C_{\textrm{eff}}$ is the total effective heat capacity. It is worth exploiting the temperature difference again, $\delta T = T_\alpha - T_\beta$, and its use leads to 
\begin{align}
T_\alpha = T + \frac{C_\beta}{C_{\textrm{eff}}} \delta T, \quad 
T_\beta = T - \frac{C_\alpha}{C_{\textrm{eff}}} \delta T. \label{eq:sub_T}
\end{align}
Substituting these relations back into the total heat flux given in Equation \eqref{eq:def_q} yields an expression for the macroscopic heat flux taking the form
\begin{align}
\mathbf{q} = -k_\alpha \nabla \left( T + \frac{C_\beta}{C_{\textrm{eff}}} \delta T \right) - k_\beta \nabla \left( T - \frac{C_\alpha}{C_{\textrm{eff}}} \delta T \right) = -k_{\textrm{eff}} \nabla T - k_{12} \nabla (\delta T). \label{eq:q_mixed}
\end{align}
Here, the homogeneous steady-state conductivity is defined as $k_{\textrm{eff}} = k_\alpha + k_\beta$; this is our static thermal conductivity, commonly used in the Fourier equation. Furthermore, we have identified the coefficient $k_{12} = \frac{k_\alpha C_\beta - k_\beta C_\alpha}{C_{\textrm{eff}}}$ as a cross-conductivity factor originating from the geometric and physical transport asymmetry existing between the two constituents.

To eliminate the temperature difference $\delta T$ from Eq.~\eqref{eq:q_mixed}, we need an independent evolution equation that describes its dynamics. We can exploit the energy balance equations, i.e., dividing Eq.~\eqref{eq:phase_a} by $C_\alpha$, and dividing Eq.~\eqref{eq:phase_b} by $C_\beta$, and subtracting the latter from the former yields
\begin{align}
\frac{\partial (\delta T)}{\partial t} &= \frac{k_\alpha}{C_\alpha} \Delta T_\alpha - \frac{k_\beta}{C_\beta} \Delta T_\beta - \left( \frac{1}{C_\alpha} + \frac{1}{C_\beta} \right) q_{\textrm{int}}. \label{eq:delta_T_raw}
\end{align}
Now, we can use the relations from Eq.~\eqref{eq:sub_T} to eliminate the phase temperatures on the right-hand side, yielding
\begin{align}
\frac{k_\alpha}{C_\alpha} \Delta T_\alpha - \frac{k_\beta}{C_\beta} \Delta T_\beta 
= \left( \frac{k_\alpha}{C_\alpha} - \frac{k_\beta}{C_\beta} \right) \Delta T + \left( \frac{k_\alpha C_\beta}{C_\alpha C_{\textrm{eff}}} + \frac{k_\beta C_\alpha}{C_\beta C_{\textrm{eff}}} \right) \Delta (\delta T). \label{eq:laplacian_expansion}
\end{align}
Simultaneously, we can substitute the definition of the non-local interfacial heat exchange $q_{\textrm{int}}$ from Eq.~\eqref{eq:nonlocal_interact} into the exchange term in Eq.~\eqref{eq:delta_T_raw}, leading to
\begin{align}
\left( \frac{1}{C_\alpha} + \frac{1}{C_\beta} \right) q_{\textrm{int}} &= \gamma \delta T + \gamma l_{\textrm{nl}}^2 \Delta (\delta T). \label{eq:exchange_expansion}
\end{align}
with $\gamma = H \left( \frac{1}{C_\alpha} + \frac{1}{C_\beta} \right)$.
Now, we can exploit the Eqs.~\eqref{eq:laplacian_expansion} and \eqref{eq:exchange_expansion} by substituting them back into the evolution equation \eqref{eq:delta_T_raw},
\begin{align}
\frac{\partial (\delta T)}{\partial t} = \left( \frac{k_\alpha}{C_\alpha} - \frac{k_\beta}{C_\beta} \right) \Delta T + \left( \frac{k_\alpha C_\beta}{C_\alpha C_{\textrm{eff}}} + \frac{k_\beta C_\alpha}{C_\beta C_{\textrm{eff}}} \right) \Delta (\delta T) - \left[ \gamma \delta T + \gamma l_{\textrm{nl}}^2 \Delta (\delta T) \right].
\end{align}
Let us define the following coefficients for simplicity,
\begin{align}
A = \frac{k_\alpha}{C_\alpha} - \frac{k_\beta}{C_\beta}, \quad
B = \frac{k_\alpha C_\beta}{C_\alpha C_{\textrm{eff}}} + \frac{k_\beta C_\alpha}{C_\beta C_{\textrm{eff}}}, \quad B^* = B - \gamma l_{\textrm{nl}}^2.
\end{align}
and after rearrangement, we obtain
\begin{align}
\frac{\partial (\delta T)}{\partial t} + \gamma \delta T - B^* \Delta (\delta T) = A \Delta T, \label{eq:delta_T_evol}
\end{align}
the evolution equation of the temperature difference.
Let us take the partial time derivative of Eq.~\eqref{eq:q_mixed},
\begin{align}
\frac{\partial \mathbf{q}}{\partial t} = -k_{\textrm{eff}} \frac{\partial (\nabla T)}{\partial t} - k_{12} \nabla \left( \frac{\partial (\delta T)}{\partial t} \right), \label{eq:q_time_deriv}
\end{align}
and eliminate the time derivative of the temperature difference $\delta T$, by substituting the evolution equation \eqref{eq:delta_T_evol} directly,
\begin{align}
\frac{\partial \mathbf{q}}{\partial t} = -k_{\textrm{eff}} \frac{\partial (\nabla T)}{\partial t} - k_{12} \nabla \left( A \Delta T - \gamma \delta T + B^* \Delta (\delta T) \right).
\end{align}
After applying the gradient operator on the bracket term,
\begin{align}
\frac{\partial \mathbf{q}}{\partial t} = -k_{\textrm{eff}} \frac{\partial (\nabla T)}{\partial t} - k_{12} A \nabla (\Delta T) + \gamma k_{12} \nabla (\delta T) - B^* \Delta \left( k_{12} \nabla (\delta T) \right), \label{eq:q_expanded_dynamic}
\end{align}
and recalling that $k_{12} \nabla (\delta T) = -\mathbf{q} - k_{\textrm{eff}} \nabla T$ is given by Eq.~\eqref{eq:q_mixed}, we obtain
\begin{align}
\frac{\partial \mathbf{q}}{\partial t} = -k_{\textrm{eff}} \frac{\partial (\nabla T)}{\partial t} - k_{12} A \nabla (\Delta T) + \gamma (-\mathbf{q} - k_{\textrm{eff}} \nabla T) - B^* \Delta (-\mathbf{q} - k_{\textrm{eff}} \nabla T).
\end{align}
We can now rearrange the equation in a way to reconstruct the common mathematical structure of non-Fourier heat equations, and introduce the macroscopic heat flux relaxation time as $\tau_q = 1/\gamma$,
\begin{align}
\tau_q \frac{\partial \mathbf{q}}{\partial t} + \mathbf{q} = -k_{\textrm{eff}} \nabla T - \tau_q k_{\textrm{eff}} \frac{\partial (\nabla T)}{\partial t} + \tau_q \left( k_{\textrm{eff}} B^* - k_{12} A \right) \nabla (\Delta T) + \tau_q B^* \Delta \mathbf{q}. \label{eq:macroeffnonfourier}
\end{align}

\subsection{Emergence of the second timescale}

Although Eq.~\eqref{eq:macroeffnonfourier} is notably similar to the Jeffreys heat equation, or to the Guyer--Krumhansl heat equation, there is a major difference, hidden in $B^*$. In the constitutive equation of the Guyer--Krumhansl model,
\begin{align}
    \tau_q \frac{\partial \mathbf{q}}{\partial t} + \mathbf{q} = - k_{\text{phonon}} \nabla T + l^2 (\Delta \mathbf{q} + 2 \nabla (\nabla \cdot \mathbf{q})), \label{eq:GK}
\end{align}
the $\mathbf{q}$ field has a non-zero rotation due to its phonon hydrodynamic background. The coefficient $l$ particularly describes the phonon mean free path, defining a typical length scale. In our case, since we assumed that each constituent obeys Fourier's law and the porosity is uniform in the isotropic medium, the macroscopic heat flux field is irrotational ($\nabla \times \mathbf{q} = \mathbf{0}$); consequently, the vector Laplacian reduces to the gradient of the divergence, $\Delta \mathbf{q} = \nabla(\nabla \cdot \mathbf{q})$. Furthermore, $l_{\textrm{nl}}$ can be treated as a continuum counterpart of $l$ describing an intrinsic length scale characteristic of the conducting medium.
In comparison with the Jeffreys heat equation,
\begin{align}
    \tau_q \frac{\partial \mathbf{q}}{\partial t} + \mathbf{q} = - k \nabla T - k \tau_T^{} \frac{\partial (\nabla T)}{\partial t},
\end{align}
however, there is a timescale independent of $\tau_q$; $\tau_T^{}$ introduces an additional dynamics beyond Fourier's law. Overall, in the following, our aim is to recover this second timescale and make the intrinsic length scale more apparent in Eq.~\eqref{eq:macroeffnonfourier}.

For this reason, we utilize the macroscopic internal energy balance, that is,
\begin{align}
    C_{\textrm{eff}} \frac{\partial T}{\partial t} = -\nabla \cdot \mathbf{q}.
\end{align}
Exploiting the irrotational property of $\mathbf{q}$, it allows us to transform the spatial non-locality entirely to the time derivative of the temperature gradient,
\begin{align}
\Delta \mathbf{q} &= \nabla \left( -C_{\textrm{eff}} \frac{\partial T}{\partial t} \right) = -C_{\textrm{eff}} \frac{\partial (\nabla T)}{\partial t}. \label{eq:vector_identity}
\end{align}
Instead of splitting the non-local effects, we substitute this identity into the $\tau_q B^* \Delta \mathbf{q}$ term, resulting in
\begin{align}
\tau_q B^* \Delta \mathbf{q} &= -\tau_q B^* C_{\textrm{eff}} \frac{\partial (\nabla T)}{\partial t}. \label{eq:elimLap}
\end{align}
We note that the non-local contribution within $B^*$ would lead to a $-l_{\textrm{nl}}^2 \Delta \mathbf{q}$ on the right-hand side, and that negative sign completely prevents the recovery of the GK equation.
Substituting Eq.~\eqref{eq:elimLap} back into Eq.~\eqref{eq:macroeffnonfourier}, we obtain a form without the Laplacian of the heat flux, yielding
\begin{align}
\tau_q \frac{\partial \mathbf{q}}{\partial t} + \mathbf{q} &= -k_{\textrm{eff}} \nabla T - \left( k_{\textrm{eff}} \tau_q + \tau_q B^* C_{\textrm{eff}} \right) \frac{\partial (\nabla T)}{\partial t} + \tau_q (k_{\textrm{eff}} B^* - k_{12} A) \nabla (\Delta T).
\end{align}
We can now expand the coefficient of the mixed derivative to define the gradient relaxation time. Recalling that $B^* = B - \gamma l_{\textrm{nl}}^2$ and $\gamma = 1 / \tau_q$, the coefficient expands as
\begin{align}
k_{\textrm{eff}} \tau_q + \tau_q B^* C_{\textrm{eff}} &= k_{\textrm{eff}} \tau_q + \tau_q B C_{\textrm{eff}} - l_{\textrm{nl}}^2 C_{\textrm{eff}} \nonumber \\
&= \tau_q \left( k_\alpha + k_\beta + \frac{k_\alpha C_\beta}{C_\alpha} + \frac{k_\beta C_\alpha}{C_\beta} \right) - l_{\textrm{nl}}^2 C_{\textrm{eff}} \nonumber \\
&= \tau_q \left[ \frac{C_{\textrm{eff}}(k_\alpha C_\beta + k_\beta C_\alpha)}{C_\alpha C_\beta} \right] - l_{\textrm{nl}}^2 C_{\textrm{eff}}.
\end{align}
Since the heat flux relaxation time is defined as $\tau_q = \frac{C_\alpha C_\beta}{H C_{\textrm{eff}}}$, the coefficient reduces to
\begin{align}
k_{\textrm{eff}} \tau_T^{} = \frac{k_\alpha C_\beta + k_\beta C_\alpha}{H} - l_{\textrm{nl}}^2 C_{\textrm{eff}} \quad \Rightarrow \quad  \tau_T^{} = \frac{k_\alpha C_\beta + k_\beta C_\alpha}{H k_{\textrm{eff}}} - \frac{l_{\textrm{nl}}^2 C_{\textrm{eff}}}{k_{\textrm{eff}}} = \tau_q \left( 1 + \frac{B^* C_{\textrm{eff}}}{k_{\textrm{eff}}} \right).
\end{align}
Here, $\tau_T^{}$ is the mathematically independent gradient relaxation time, characterizing the additional timescale beyond Fourier. It is important to highlight that the non-local effect decreases $\tau_T^{}$, and thus stands as a strong physical limit to preserve the thermodynamic admissibility of the macroscopic equation ($\tau_T^{}>0$) by
\begin{align}
    l_{\textrm{nl}}^2 \leq \frac{k_\alpha C_\beta + k_\beta C_\alpha}{H C_{\textrm{eff}}}. \label{eq:48}
\end{align}
In order to simplify the notations, we also introduce $\mathcal{D} = \tau_q (k_{\textrm{eff}} B^* - k_{12} A)$, and that coefficient can be simplified after a straightforward algebraic manipulation,
\begin{align}
    \mathcal{D} = \frac{k_\alpha k_\beta}{H} - k_{\textrm{eff}} l_{\textrm{nl}}^2, \label{eq:D}
\end{align}
and thus we arrive at the final form of the effective macroscopic non-Fourier heat equation,
\begin{align}
\tau_q \frac{\partial \mathbf{q}}{\partial t} + \mathbf{q} &= -k_{\textrm{eff}} \nabla T - k_{\textrm{eff}} \tau_T^{} \frac{\partial (\nabla T)}{\partial t} + \mathcal{D} \nabla (\Delta T). \label{eq:macroeffnonfou2}
\end{align}
Interestingly, the heterogeneous structure with two heat conduction channels leads directly to a non-Fourier heat equation after upscaling to a macroscopic length scale. This effective model successfully reconstructs a Jeffreys-type equation with an additional term. However, it is important to note that this macroscopic effective model cannot recall the Guyer--Krumhansl equation, as it lacks a positive $\Delta \mathbf{q}$ term on the right-hand side. The GK equation originates from phonon hydrodynamics, where the heat flux possesses a rotational component, and the non-local Laplacian term is directly connected to the phonon mean free path.

\section{Notes on thermodynamic admissibility}

In phenomenological irreversible thermodynamic theories, the thermodynamic admissibility of non-Fourier heat conduction is enforced by postulating a generalized macroscopic entropy function $s(\mathbf{q}, T)$. The coefficients of the resulting transport equations are subsequently constrained to ensure positive macroscopic entropy production ($\sigma_{macro} \ge 0$). Unlike phenomenological approaches, within the LTNE volume averaging framework, thermodynamic admissibility is not postulated; it is a consequence of the underlying microstructural evolution equations. However, the discovered $\nabla \Delta T$ term necessitates deeper insight into thermodynamic compatibility. In order to demonstrate that the derived macroscopic Jeffreys-type equation satisfies the second law of thermodynamics, we formulate the Clausius--Duhem inequality directly from the phase-level entropy balances.

\subsection{Phase-level entropy production}
Assuming classical Fourier conduction within the stationary microscopic phases, the local entropy production is unconditionally non-negative. Upon volume averaging, the total macroscopic specific entropy production rate ($\sigma_{macro}$) is the sum of the entropy generated by heat conduction within each phase, and the thermal resistance at the microscopic interface,
\begin{equation}
    \sigma_{macro} = \mathbf{q}_\alpha \cdot \nabla \left( \frac{1}{T_\alpha} \right) + \mathbf{q}_\beta \cdot \nabla \left( \frac{1}{T_\beta} \right) + q_{\textrm{int}} \left( \frac{1}{T_\beta} - \frac{1}{T_\alpha} \right) \ge 0. \label{eq:entropy_base}
\end{equation}
For linear transport processes near equilibrium, we can apply linearization $\nabla (1/T_i) \approx -\nabla T_i / T_0^2$, where $T_0$ is a constant reference temperature. By utilizing local algebraic interfacial exchange ($q_{\textrm{int}} \approx H(T_\alpha - T_\beta)$) and substituting the Fourier law ($\mathbf{q}_\alpha = -k_\alpha \nabla T_\alpha$ and $\mathbf{q}_\beta = -k_\beta \nabla T_\beta$), we obtain a strictly positive quadratic form,
\begin{equation}
    T_0^2 \sigma_{macro} = k_\alpha (\nabla T_\alpha)^2 + k_\beta (\nabla T_\beta)^2 + H (T_\alpha - T_\beta)^2 \ge 0. \label{eq:entropy_linear}
\end{equation}

In order to evaluate the thermodynamic constraints introduced by the non-local term in $q_{\textrm{int}}$, we apply the identity $(T_\alpha - T_\beta) \Delta (T_\alpha - T_\beta) = \nabla \cdot \left[ (T_\alpha - T_\beta) \nabla (T_\alpha - T_\beta) \right] - \left( \nabla (T_\alpha - T_\beta) \right)^2$. The divergence term represents an extra entropy flux originating from the microstructural interface. Extracting the dissipative components provides the entropy production,
\begin{equation}
    T_0^2 \sigma_{macro}^{\textrm{eff}} = k_\alpha (\nabla T_\alpha)^2 + k_\beta (\nabla T_\beta)^2 + H (T_\alpha - T_\beta)^2 - H l_{\textrm{nl}}^2 \left( \nabla (T_\alpha - T_\beta) \right)^2 \ge 0. \label{eq:entropy_effective}
\end{equation}
Notably, while the local exchange ($H$) generates entropy, the non-local term ($-H l_{\textrm{nl}}^2$) acts conversely, decreasing the local entropy production. Consequently, this necessitates a strict upper bound on the non-local length scale $l_{\textrm{nl}}$ to ensure that the thermodynamic continuum remains asymptotically stable.

\subsection{Determining the upper bound of $l_{\textrm{nl}}$}

Based on the elimination of phase temperatures in Section 3, it is worth rewriting the entropy production using the variables of $T$ and $\delta T$. Let us realize that Eq.~\eqref{eq:entropy_effective} can be rewritten in a matrix form,
\begin{equation}
    T_0^2 \sigma_{macro}^{\textrm{eff}} = \begin{bmatrix} \nabla T & \nabla (\delta T) \end{bmatrix} \underbrace{\begin{bmatrix} k_{\textrm{eff}} & k_{12} \\ k_{12} & k_{\textrm{int}} - H l_{\textrm{nl}}^2 \end{bmatrix}}_{\mathbf{M}} \begin{bmatrix} \nabla T \\ \nabla (\delta T) \end{bmatrix} + H (\delta T)^2 \ge 0,
\end{equation}
using the following coefficients,
\begin{equation}
    k_{\textrm{eff}} = k_\alpha + k_\beta, \quad k_{12} = \frac{k_\alpha C_\beta - k_\beta C_\alpha}{C_{\textrm{eff}}}, \quad \quad k_{\textrm{int}} = \frac{k_\alpha C_\beta^2 + k_\beta C_\alpha^2}{C_{\textrm{eff}}^2},
\end{equation}
with the variables of $\nabla T$ and $\delta T$.
To ensure the positive definite entropy production, the conductivity matrix $\mathbf{M}$ must be positive definite. Because $k_{\textrm{eff}} > 0$, Sylvester's criterion requires the determinant of the matrix to be positive definite,
\begin{equation}
    \text{det}(\mathbf{M}) = k_{\textrm{eff}} (k_{\textrm{int}} - H l_{\textrm{nl}}^2) - k_{12}^2 > 0.
\end{equation}
Evaluating the determinant, we can observe the separation of the local and non-local contributions,
\begin{equation}
    \text{det}(\mathbf{M}) = (k_{\textrm{eff}} k_{\textrm{int}} - k_{12}^2) - k_{\textrm{eff}} H l_{\textrm{nl}}^2.
\end{equation}
The term $(k_{\textrm{eff}} k_{\textrm{int}} - k_{12}^2)$ reduces to $k_\alpha k_\beta$. Therefore, the non-local determinant is simplified to
\begin{equation}
    \text{det}(\mathbf{M}) = k_\alpha k_\beta - k_{\textrm{eff}} H l_{\textrm{nl}}^2 > 0. \label{eq:determinant_reduction_nonlocal}
\end{equation}

Recall that the macroscopic non-local coefficient of the higher-order spatial behavior reduces to $\mathcal{D} = \frac{k_\alpha k_\beta}{H} - k_{\textrm{eff}} l_{\textrm{nl}}^2$.
After rearranging Eq.~\eqref{eq:determinant_reduction_nonlocal}, one immediately observes that the determinant of the conductivity matrix $\mathbf{M}$ is proportional to the macroscopic spatial non-local coefficient,
\begin{equation}
    \text{det}(\mathbf{M}) = H \left( \frac{k_\alpha k_\beta}{H} - k_{\textrm{eff}} l_{\textrm{nl}}^2 \right) = H \mathcal{D} > 0.
\end{equation}
The condition $\text{det}(\mathbf{M}) > 0$ ensures a thermodynamic upper limit on the length scale,
\begin{equation}
    0<l_{\textrm{nl}}^2 < \frac{k_\alpha k_\beta}{k_{\textrm{eff}} H} \quad \Rightarrow \quad l_{\textrm{nl}} < \sqrt{\frac{k_\alpha k_\beta}{k_{\textrm{eff}} H}},
\end{equation}
which is, interestingly, a stricter constraint than requiring $\tau_T^{}>0$ by Eq.~\eqref{eq:48}.

\section{Model reduction}

Let us investigate the one-dimensional version of the extended Jeffreys heat equation in order to study the order of magnitude of each term, particularly the third-order derivative in
\begin{align}
    \tau_q \frac{\partial q}{\partial t} + q &= -k_{\textrm{eff}} \frac{\partial T}{\partial x} - k_{\textrm{eff}} \tau_T^{} \frac{\partial^2 T}{\partial t \partial x} + \mathcal{D} \frac{\partial^3 T}{\partial x^3}.
\end{align}

Based on the typical parameters for porous rocks in a heat pulse experiment (temperature rise $\delta T_{\textrm{mac}} \approx 4$ K, sample thickness $L \approx 2.5$ mm), we establish the following reference quantities based on \cite{FehEtal21, FehKov24} and the known origin of the macroscopic transport parameters derived in Sec.~3,
\begin{itemize}
    \item effective volumetric heat capacity: $C_{\textrm{eff}} \approx 2 \times 10^6 \text{ J/(m}^3\text{K)}$;
    \item Fourier (static) thermal conductivity: $k_{\textrm{eff}} \approx 2 \text{ W/(m K)}$;
    \item Fourier (static) diffusivity: $a_{\textrm{eff}} = k_{\textrm{eff}} / C_{\textrm{eff}} \approx 10^{-6} \text{ m}^2/\text{s}$;
    \item heat flux relaxation time: $\tau_q \approx 0.5 \text{ s}$;
    \item dynamic-to-static conductivity ratio: $k_{\textrm{dyn}}/k_{\textrm{stat}} \approx 1.4 \implies \tau_T^{} \approx 0.7 \text{ s}$;
    \item maximum spatial non-locality coefficient: $\mathcal{D}_{\textrm{max}} \approx 0.49 \times 10^{-6} \text{ W m}/\text{K}$ (assuming $a_\alpha \approx a_\beta$ to maximize the product).
\end{itemize}

In such an experiment, the transport process can be divided into two distinct time intervals, each requiring a different characteristic scale for accurate order-of-magnitude estimates.

\subsection{Early transient behavior}

During the steep rise of the heat pulse ($t \approx \tau_q = 0.5$ s), the characteristic spatial scale is not the total sample thickness. Instead, we must use the {thermal penetration depth}, denoted as $x_c$. 
At early times ($t \le \tau_q$), the thermal front has not yet traversed the entire sample length $L$. The region of steep temperature gradients is strictly limited to a thin layer near the excitation boundary. The spatial extent of this region is governed by the dynamic thermal penetration depth, $x_c = \sqrt{a_{\textrm{dyn}} \tau_q}$, i.e., using the dynamic thermal conductivity \cite{FehKov24}. If we were to use the sample thickness $L$ to approximate the spatial derivatives, we would artificially distort the temperature gradients ($\partial_x T$, and especially $\partial_{xxx} T$). This would lead to a significant underestimation of the higher-order spatial derivatives.

\noindent Using the appropriate early-timescales:
\begin{itemize}
    \item characteristic time: $t_c = \tau_q = 0.5 \text{ s}$;
    \item characteristic length: $x_c = \sqrt{a_{\textrm{dyn}} \tau_q} \approx 0.84 \times 10^{-3} \text{ m}$;
    \item characteristic heat flux: $q_c \approx k_{\textrm{eff}} \delta T_{\textrm{mac}} / x_c \approx 9520 \text{ W/m}^2$.
\end{itemize}
Evaluating the terms of the generalized Jeffreys equation yields the results presented in Table \ref{tab:1}. The third-order spatial derivative $\mathcal{D} \partial_{xxx} T$ represents approximately 25-35\% of the leading transport terms. Interestingly, such a third-order term is completely hidden in both extended irreversible thermodynamics and in the internal variable theory. Consequently, this estimate suggests that the re-evaluation of the existing experiments would be preferable in order to confirm the magnitude of $\mathcal{D}$ or the omission of such a term.

\begin{table}[h]
\centering
\caption{Order of magnitude of the terms during the early transient ($t \approx 0.5$ s).}
\renewcommand{\arraystretch}{1.5}
\begin{tabular}{llc}
{Term} & {Scaling} & {Magnitude (W/m$^2$)} \\
Heat flux rate ($\tau_q \partial_t q$) & $\tau_q \cdot (q_c / t_c)$ & $\sim 9520$ \\
Heat flux ($q$) & $q_c$ & $\sim 9520$ \\
Fourier gradient ($k_{\textrm{eff}} \partial_x T$) & $k_{\textrm{eff}} \cdot (\delta T_{\textrm{mac}} / x_c)$ & $\sim 9520$ \\
Mixed derivative ($k_{\textrm{eff}} \tau_T^{} \partial_{xt} T$) & $k_{\textrm{eff}} \tau_T^{} \cdot (\delta T_{\textrm{mac}} / (x_c t_c))$ & $\sim 13330$ \\
Spatial third-order ($\mathcal{D} \partial_{xxx} T$) & $\mathcal{D}_{\textrm{max}} \cdot (\delta T_{\textrm{mac}} / x_c^3)$ & $\sim 3310$ \\
\end{tabular} \label{tab:1}
\end{table}

\subsection{Late transient behavior}

As time progresses much beyond the relaxation time ($t \gg \tau_q$), the heat pulse broadens and penetrates the entire sample, and the system approaches thermal equilibrium. 
In this regime, the physical boundaries of the sample dictate the transport scales:
\begin{itemize}
    \item characteristic length: $x_c = L \approx 2.5 \times 10^{-3} \text{ m}$;
    \item characteristic time: $t_c = L^2 / a_{\textrm{eff}} \approx 6.25 \text{ s}$;
    \item characteristic heat flux: $q_c \approx k_{\textrm{eff}} \delta T_{\textrm{mac}} / L \approx 3200 \text{ W/m}^2$;
\end{itemize}
Evaluating the terms for the late transient yields the results presented in Table \ref{tab:2}. During diffusion and the subsequent cooling phase, classical Fourier transport completely dominates energy transfer. The non-local spatial term $\mathcal{D} \partial_{xxx} T$ notably decreases to less than 4\% of the primary terms, indicating that it can be neglected in the long-time asymptotic analysis.

\begin{table}[h]
\centering
\caption{Order of magnitude of the terms during the late transient ($t \gg \tau_q$).}
\renewcommand{\arraystretch}{1.5}
\begin{tabular}{llc}

{Term} & {Scaling} & {Magnitude (W/m$^2$)} \\

Heat flux rate ($\tau_q \partial_t q$) & $\tau_q \cdot (q_c / t_c)$ & $\sim 256$ \\
Heat flux ($q$) & $q_c$ & $\sim 3200$ \\
Fourier gradient ($k_{\textrm{eff}} \partial_x T$) & $k_{\textrm{eff}} \cdot (\delta T_{\textrm{mac}} / x_c)$ & $\sim 3200$ \\
Mixed derivative ($k_{\textrm{eff}} \tau_T^{} \partial_{xt} T$) & $k_{\textrm{eff}} \tau_T^{} \cdot (\delta T_{\textrm{mac}} / (x_c t_c))$ & $\sim 358$ \\
Spatial third-order ($\mathcal{D} \partial_{xxx} T$) & $\mathcal{D}_{\textrm{max}} \cdot (\delta T_{\textrm{mac}} / x_c^3)$ & $\sim 125$ \\
\end{tabular} \label{tab:2}
\end{table}

\section{Comparison with the 2T model}

It is necessary to compare the derived macroscopic effective model with the classical two-temperature (2T) model. Both models attempt to capture deviations in the thermal response due to microstructural heterogeneity. Despite their mathematical similarities, there are essential differences as well.

Beginning with their similarities and studying the $T$-representation of both models. We assume local heat exchange ($q_{\textrm{int}} = H(T_\alpha - T_\beta)$), which  neglects finite-boundary non-localities ($l_{\textrm{nl}} = 0$), and thus the macroscopic evolution equations of the 2T model are
\begin{align}
    C_\alpha \frac{\partial T_\alpha}{\partial t} &= k_\alpha \Delta T_\alpha - H(T_\alpha - T_\beta), \\
    C_\beta \frac{\partial T_\beta}{\partial t} &= k_\beta \Delta T_\beta + H(T_\alpha - T_\beta).
\end{align}
By applying the macroscopic mixture temperature $T = (C_\alpha T_\alpha + C_\beta T_\beta)/C_{\textrm{eff}}$, and eliminating both phase temperatures, the system reduces to 
\begin{equation}
    \frac{C_\alpha C_\beta}{H C_{\textrm{eff}}} \frac{\partial^2 T}{\partial t^2} + \frac{\partial T}{\partial t} = \frac{k_{\textrm{eff}}}{C_{\textrm{eff}}} \Delta T + \frac{k_\alpha C_\beta + k_\beta C_\alpha}{H C_{\textrm{eff}}} \frac{\partial (\Delta T)}{\partial t} - \frac{k_\alpha k_\beta}{H C_{\textrm{eff}}} \Delta^2 T, \label{eq:exact_T}
\end{equation}
which is the $T$-representation of the 2T model.
In order to have a direct comparison, we take the divergence of the macroscopic effective flux equation (Eq.~\ref{eq:macroeffnonfou2}) and substitute the macroscopic energy balance ($\nabla \cdot \mathbf{q} = -C_{\textrm{eff}} \frac{\partial T}{\partial t}$), yielding
\begin{equation}
    \tau_q \frac{\partial^2 T}{\partial t^2} + \frac{\partial T}{\partial t} = a_{\textrm{eff}} \Delta T + \tau_T^{} a_{\textrm{eff}} \frac{\partial (\Delta T)}{\partial t} - \frac{\mathcal{D}}{C_{\textrm{eff}}} \Delta^2 T. \label{eq:divergence_macro}
\end{equation}

Comparing Eq.~\eqref{eq:exact_T} and Eq.~\eqref{eq:divergence_macro} demonstrates that the classical 2T model is structurally analogous to the derived effective non-Fourier model, taking the form of a generalized Jeffreys equation with an additional spatial derivative. However, there are crucial differences as well.

It holds for both models that eliminating any (macroscopic) field variables is unfavorable, as it hides the model's thermodynamic structure. Consequently, it is remarkably difficult to handle non-homogeneous initial states or complex boundary conditions using the $T$-representation. The separate use of the evolution equations for $\mathbf{q}$ and $T$ reveals the major difference in the origin of timescales. In the 2T model, the static and dynamic timescales are hidden, especially on the constitutive level. Treating $T_\alpha$ and $T_\beta$ as separate field variables, it is not clear which one, and in what way, determines the static and dynamic timescales. In the case of the present generalized Jeffreys equation, the ratio of $\tau_T^{}/\tau_q$ uniquely determines the contribution of constituents to both timescales.

Although both models lead to the same ratio of the static and dynamic timescales due to their identical $T$-representation, the solution methods and the definitions for initial and boundary conditions can significantly differ due to the evolution equation of $\mathbf{q}$. In the 2T model, one can define separate initial and boundary conditions for each phase, but this leads to significant uncertainties. For example, when a heat pulse is applied to a metal foam, it is physically ambiguous how to partition the incoming boundary heat flux mathematically between the solid matrix and the fluid phases in the 2T model. Because the generalized Jeffreys equation models the effective continuum heat flux directly, it avoids this phase-partitioning ambiguity due to the upscaling of the heterogeneous microstructure.

Additionally, by separating the length scales, further non-local extensions are possible.
In the volume-averaged model, recognizing that thermal exchange occurs over a spatially distributed interface introduces the non-local length scale $l_{\textrm{nl}}$. However, we note that despite the mathematical structure of the generalized Jeffreys equation, it is also not able to model wave propagation -- analogously to the 2T model.

\section{The ratio of timescales and the emergence of scale dependence}

To quantify the deviation from classical Fourier conduction, we define the over-diffusion coefficient, $\mathcal{R}$, as
\begin{equation}
    \mathcal{R} = \frac{\tau_T^{}}{\tau_q},
\end{equation}
similarly to the paper \cite{Botetal16}.
In classical Fourier theory, these timescales are presumed identical ($\tau_T^{} = \tau_q$), implying a resonance condition of $\mathcal{R} = 1$. We note that $\tau_T^{}=\tau_q=0$ is often considered during the model reduction, but in this case, these coefficients cannot be zero, and the appearance of additional timescales is more natural in this way.
We can substitute the definitions for both relaxation times, revealing the component-wise contribution,
\begin{equation}
    \mathcal{R} = \frac{ \frac{k_\alpha C_\beta + k_\beta C_\alpha}{H k_{\textrm{eff}}} - \frac{l_{\textrm{nl}}^2 C_{\textrm{eff}}}{k_{\textrm{eff}}} }{ \frac{C_\alpha C_\beta}{H C_{\textrm{eff}}} } = \frac{(k_\alpha C_\beta + k_\beta C_\alpha) C_{\textrm{eff}}}{k_{\textrm{eff}} C_\alpha C_\beta} - \frac{l_{\textrm{nl}}^2 H C_{\textrm{eff}}^2}{k_{\textrm{eff}} C_\alpha C_\beta}.
\end{equation}
By expanding the first term, we isolate the scale-independent, localized contribution ($\mathcal{R}^*$) from the finite-boundary non-local contribution,
\begin{equation}
    \mathcal{R} = \underbrace{ 1 + \frac{k_\alpha C_\beta^2 + k_\beta C_\alpha^2}{C_\alpha C_\beta k_{\textrm{eff}}} }_{\mathcal{R}^*} - \frac{l_{\textrm{nl}}^2 H C_{\textrm{eff}}^2}{k_{\textrm{eff}} C_\alpha C_\beta}.
\end{equation}
Recalling the definition of the interaction conductivity $k_{\textrm{int}} = (k_\alpha C_\beta^2 + k_\beta C_\alpha^2)/C_{\textrm{eff}}^2$ introduced during the entropy production analysis, the over-diffusion coefficient reduces to
\begin{equation}
    \mathcal{R} = 1 + \frac{C_{\textrm{eff}}^2}{C_\alpha C_\beta k_{\textrm{eff}}} \left( k_{\textrm{int}} - H l_{\textrm{nl}}^2 \right). \label{eq:R_calc_def}
\end{equation}

\subsection{Recovering Fourier resonance}

Equation \eqref{eq:R_calc_def} provides insight regarding the origin of over-diffusion in heterogeneous media. If the macroscopic boundaries are assumed infinite or the interfacial exchange is local ($l_{\textrm{nl}} = 0$), the ratio reduces to $\mathcal{R}^*$. Because the thermophysical properties are strictly positive, the additive term is positive ($k_{\textrm{int}} > 0 \Rightarrow \mathcal{R}^* > 1$). Therefore, the model is inevitably over-diffusive, explaining why this region is exclusively observed in our experiments, as in \cite{FehEtal21, FehEtal24b}. However, it is central to highlight that this does not mean that over-diffusion can be easily observed in any heterogeneous material. The over-diffusion coefficient characterizes only the material timescales, but direct observation of over-diffusion also requires an appropriate set of boundary conditions \cite{Kov24}. Furthermore, for very fast excitations (i.e., on the dynamic timescale), only the dynamic thermal conductivity $k_{\textrm{eff}}\mathcal{R}$ is dominant. It underlines again that both components of the heterogeneous material are needed for over-diffusion and for the realization of the dynamic timescale. This is not that apparent in the 2T model.

However, the inclusion of the non-local length scale $l_{\textrm{nl}}$ might effectively lead to Fourier resonance ($\mathcal{R} = 1$), but only under the strict condition that the non-local interfacial thermal exchange exactly balances the internal phase properties,
\begin{equation}
    \mathcal{R} = 1 \quad \iff \quad l_{\textrm{nl}}^2 = \frac{k_{\textrm{int}}}{H}.
\end{equation}
This is a purely mathematical requirement rather than a universally applicable condition for real heterogeneous materials; therefore, we do not further exploit the non-local coefficient.
In most real materials with highly distinct phases (e.g., metal foams, porous rocks), the asymmetry dominates, resulting in the experimentally observed over-diffusive regime ($\tau_T^{} > \tau_q$).

\subsection{Emergence of the apparent scale-dependence $\mathcal{R}_{\textrm{app}}(L_{\textrm{sample}})$}

In order to reveal any scale dependence, we must assume that the sample becomes comparable to the RVE length scale, and thus might fall into the sub-RVE domain.
The sample is bounded by a finite thickness, $L\approx L_{\textrm{sample}}< L_{\textrm{mac}}$, with heat propagating along the $x$-axis.
Following the analytical solution of the Jeffreys heat equation from \cite{FehKov24} to obtain the transient temperature field $T(x,t)$ within adiabatic boundaries after the heat pulse, $T(x,t)$ is described by a Fourier series of exponentially decaying spatial eigenmodes. Because higher-order modes decay rapidly, the dynamic thermal response right after the heat pulse is dominated by the first eigenmode,
\begin{equation}
    T(x,t) - T_{\infty} \approx A_1 \exp\left(-\frac{t}{\tau_1}\right) \cos\left( \frac{\pi x}{L_{\textrm{sample}}} \right).
\end{equation}
Applying the Laplacian leads to
\begin{equation}
    \Delta T \approx -\frac{\pi^2}{L_{\textrm{sample}}^2} (T - T_{\infty}),
\end{equation}
and repeating it again, it suggests the scaling
\begin{equation}
    \Delta^2 T \approx -\frac{\pi^2}{L_{\textrm{sample}}^2} \Delta T. \label{eq:modal_scaling}
\end{equation}
Substituting this modal mapping into the $T$-representation of the generalized Jeffreys equation,
\begin{equation}
    \tau_q \frac{\partial^2 T}{\partial t^2} + \frac{\partial T}{\partial t} = \underbrace{\left( a_{\textrm{eff}} + \frac{\mathcal{D}}{C_{\textrm{eff}}} \frac{\pi^2}{L_{\textrm{sample}}^2} \right)}_{a_{\textrm{app}}} \Delta T + \tau_q \underbrace{\left(\frac{\tau_T^{} a_{\textrm{eff}}}{\tau_q}\right)}_{a_{\textrm{dyn,app}}} \frac{\partial (\Delta T)}{\partial t},
\end{equation}
in which we replaced the fourth-order spatial derivative by Eq.~\eqref{eq:modal_scaling}.
The apparent gradient relaxation time becomes $\tau_{T,\textrm{app}} = \tau_T^{} (a_{\textrm{eff}}/a_{\textrm{app}})$, and this could be the origin of observing size-dependent behavior for rocks in \cite{FehEtal21}.  
Consequently, the apparent over-diffusive coefficient measured in finite sub-RVE samples, $\mathcal{R}_{\textrm{app}}(L_{\textrm{sample}}) = \tau_{T,\textrm{app}} / \tau_q$ reads
\begin{equation}
    \mathcal{R}_{\textrm{app}}(L_{\textrm{sample}}) = \mathcal{R} \left( \frac{a_{\textrm{eff}}}{a_{\textrm{eff}} + \frac{\mathcal{D}}{C_{\textrm{eff}}} \frac{\pi^2}{L_{\textrm{sample}}^2}} \right) = \mathcal{R} \left[ 1 + \frac{\mathcal{D} \pi^2}{k_{\textrm{eff}} L_{\textrm{sample}}^2} \right]^{-1}, \label{eq:R_ratio_exact}
\end{equation}
in which the correction term can be further expressed using Eq.~\eqref{eq:D} (with $l_{\textrm{nl}}=0$) and $k_{\textrm{eff}} = k_\alpha + k_\beta$, yielding
\begin{align}
    \mathcal{R}_{\textrm{app}}(L_{\textrm{sample}}) = \mathcal{R} \left[ 1 + \frac{\mathcal{D} \pi^2}{k_{\textrm{eff}} L_{\textrm{sample}}^2} \right]^{-1} = \mathcal{R} \left[ 1 +  \frac{k_\alpha k_\beta}{H} \frac{ \pi^2}{(k_\alpha + k_\beta) L_{\textrm{sample}}^2} \right]^{-1}.
\end{align}
The volumetric heat transfer coefficient is usually very large, on the order of $10^{6}$ W/(m$^3$ K); thus, the correction term is most relevant for thin samples.

\section{Experimental validations}

\begin{figure}
    \centering
    \includegraphics[width=0.95\linewidth]{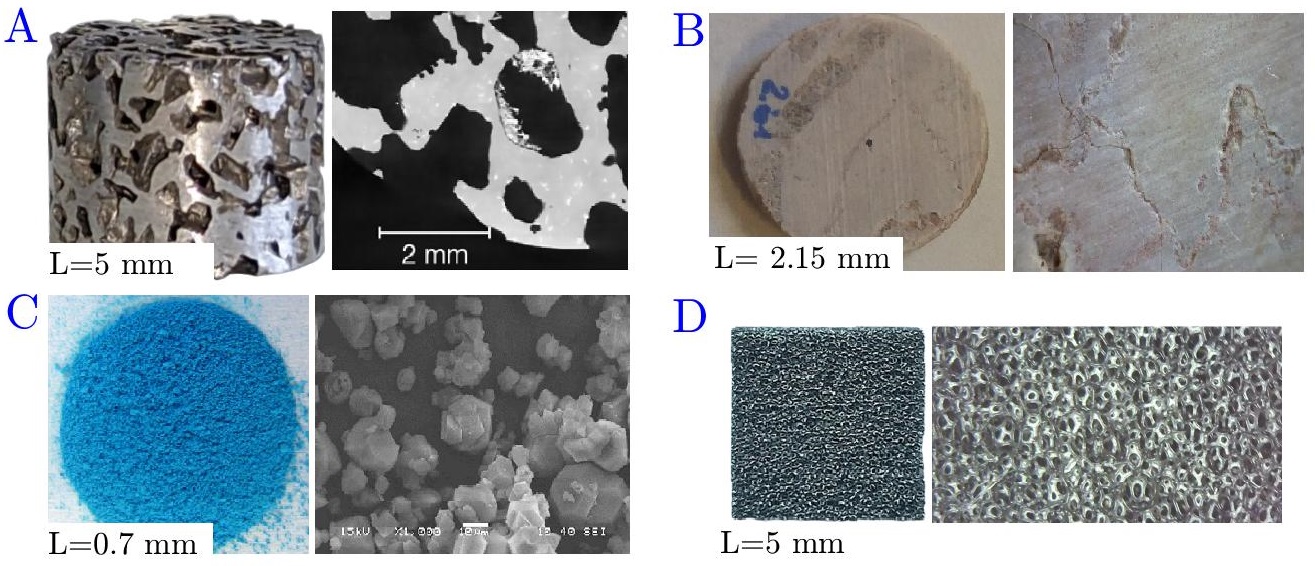}
    \caption{Typical heterogeneous structure exhibiting effective non-Fourier behavior. A) Metal foam, investigated by Lunev et al.~\cite{LunEtal22, LauererLunev23}. B) Rock sample from \cite{FehEtal21}. C) Metal-organic framework from \cite{GalEtal24}. D) Highly porous carbon foam from \cite{FehEtal24}. The figure is taken from \cite{Kov24}, and $L$ denotes the sample thickness.}
    \label{fig:2}
\end{figure}

In recent years, we have conducted numerous heat conduction experiments on various foams, metal-organic frameworks, and rocks \cite{Botetal16, FehEtal24, FehEtal21, FehEtal24b, GalEtal24}. The tested samples are often highly heterogeneous, including high porosity and multi-component structures on various length scales. Figure \ref{fig:2} provides a brief overview of the variety of materials tested. Most of these heterogeneous materials exhibited non-Fourier behavior at length scales of 1--3 mm. Although the thickness is strictly limited by the flash apparatus, this limitation enabled the observation of multiple timescales \cite{Kov24}. In the following, we study typical heterogeneous materials using the present framework.

\subsubsection*{Case 1: Air-filled aluminum foam}
Beginning with an open-cell aluminum foam, and we assign $\alpha=s$ for the solid, and $\beta=f$ for the fluid phase, i.e., $\varepsilon_s = 0.10$, $\varepsilon_f = 0.90$, pore diameter $d_p = 0.002 \, \mathrm{m}$. We utilize Lemlich's approximation for a statistically isotropic Kelvin cell \cite{Lemlich78, KrishEtal08}, where one-third of the solid aligns with any arbitrary one-dimensional macroscopic gradient, yielding the effective phase conductivities: $k_s = \frac{1}{3} \varepsilon_s k_{s,m}$ and $k_f = \frac{1}{3} \varepsilon_f k_{f,m}$, where subscript $m$ denotes the microscopic material property. Assuming standard thermophysical constants for aluminum ($k_{s,m} = 200 \, \mathrm{W/(m \ K)}$, $\rho_s c_{s} = 2.43 \cdot 10^6 \, \mathrm{J/(m^3 \ K)}$) and stagnant air ($k_{f,m} = 0.026 \, \mathrm{W/(m \ K)}$, $\rho_f c_{f} = 1206 \, \mathrm{J/(m^3 \ K)}$), we obtain the effective capacities for the solid $C_s = \varepsilon_s \rho_s c_{ps} = 243 \, \mathrm{kJ/(m^3 \ K)}$, and the fluid phase, $C_f = \varepsilon_f \rho_f c_{pf} = 1085 \, \mathrm{J/(m^3 \ K)}$. The conductivities are $k_s = 6.667 \, \mathrm{W/(m \ K)}$ and $k_f = 0.008 \, \mathrm{W/(m \ K)}$. 

In order to estimate the volumetric exchange coefficient, we follow the studies of \cite{LuEtal98, CalmMah00}. We can assume that $H = a_v h$, where the specific surface area is approximated as $a_v \approx 14/d_p$ (with a pore diameter $d_p \approx 0.002$ m), and the heat transfer coefficient is $h = Nu \cdot k_{f,m} / d_p$ (with $Nu \approx 6.0$), resulting in
\begin{equation}
    H_{air} = \frac{84 k_{f,m}}{d_p^2} = 546 \, \mathrm{kW/(m^3 \ K)}.
\end{equation}
After using the definitions for $\tau_q$ and $k_{\textrm{eff}}$, we find $\tau_q \approx 2.0 \, \mathrm{ms}$, and the static thermal conductivity is $k_{\textrm{eff}} = 6.675 \, \mathrm{W/(m \ K)}$. Assuming the absence of non-local heat transfer ($l_{\textrm{nl}}=0$, hence $\mathcal R = \mathcal R^*$), the over-diffusion coefficient becomes $\mathcal{R} \approx 1.27$. This can be considered as a moderate over-diffusion coefficient, since the high porosity of the metal foam significantly decreases its effective thermal conductivity. 

To investigate the macroscopic boundary effects, we evaluate the size-dependent apparent over-diffusion coefficient $\mathcal{R}_{\textrm{app}}(L_{\textrm{sample}})$ for finite samples. Because the fluid phase (air) has a low thermal conductivity, the product $k_s k_f$ is remarkably small, resulting in a low maximum spatial non-locality coefficient of $\mathcal{D} \approx 9.5 \cdot 10^{-8} \, \mathrm{W \ m / K}$. Consequently, the higher-order spatial correction $\frac{\mathcal{D} \pi^2}{k_{\textrm{eff}} L_{\textrm{sample}}^2}$ exerts a noticeable influence only for very thin samples. Table \ref{tab:air_foam} summarizes $\mathcal{R}_{\textrm{app}}$ for typical macroscopic lengths, demonstrating that the non-local boundary effect becomes essentially negligible for $L_{\textrm{sample}} \ge 5 \, \mathrm{mm}$.

\begin{table}[h]
\centering
\caption{Apparent over-diffusion coefficient $\mathcal{R}_{\textrm{app}}$ for an air-filled aluminum foam at various macroscopic sample thicknesses.}
\renewcommand{\arraystretch}{1.5}
\begin{tabular}{l c}
$L_{\textrm{sample}}$ (mm) & $\mathcal{R}_{\textrm{app}}$ \\ \hline
2 & 1.22 \\
5 & 1.26 \\
20 & 1.27 \\
100 & 1.27 \\ \hline
\end{tabular} \label{tab:air_foam}
\end{table}

\subsubsection*{Case 2: Water-filled aluminum foam}
For innovative heat exchanger designs \cite{HanEtal12, Booms03}, it can be relevant if the cells are filled with water. 
Assuming the same aluminum matrix, but now with water ($k_{f,m} = 0.6 \, \mathrm{W/(m \ K)}$, $\rho_f c_{f} = 4.18 \cdot 10^6 \, \mathrm{J/(m^3 \ K)}$) notably changes the thermal asymmetry. The fluid heat capacity increases to $C_f = 3.762 \cdot 10^6 \, \mathrm{J/(m^3 \ K)}$, and the effective fluid conductivity becomes $k_f = 0.18 \, \mathrm{W/(m \ K)}$. The higher fluid conductivity significantly amplifies the pore-scale thermal exchange, yielding $H_{water} = 12.6 \cdot 10^6 \mathrm{W/(m^3 \ K)}$, and resulting in $\tau_q\approx 18.1 \, \mathrm{ms}$. The static thermal conductivity increases slightly to $k_{\textrm{eff}} = 6.847 \, \mathrm{W/(m \ K)}$. However, evaluating the over-diffusion coefficient reveals $\mathcal{R} \approx 16.1$, thanks to the much higher fluid heat capacity. Experiments in this direction have not yet been performed to the author's knowledge, but this scale-dependent behavior could be a central structural property in heat exchangers, especially in heat storage systems with significant transients. On the other hand, since $\tau_q$ is about 20 ms, this timescale is completely negligible in usual industrial heat exchangers, but can be interesting for special applications, e.g., in the space industry.  

To evaluate the macroscopic boundary effects, we analyze the size-dependent apparent over-diffusion coefficient, $\mathcal{R}_{\textrm{app}}(L_{\textrm{sample}})$. Interestingly, based on the geometric relations, the fluid's microscopic thermal conductivity cancels out when calculating the spatial non-locality coefficient. As a result, $\mathcal{D}$ is identical to the air-filled case at $\approx 9.5 \cdot 10^{-8} \, \mathrm{W \ m / K}$. Because $k_{\textrm{eff}}$ only increases marginally with water, the magnitude of the higher-order spatial correction is very similar, primarily restricting the over-diffusion coefficient in samples thinner than $5 \, \mathrm{mm}$. Table \ref{tab:water_foam} illustrates this scaling, demonstrating how the finite boundary effects suppress $\mathcal{R}_{\textrm{app}}$ before it converges to its large bulk value.

\begin{table}[h]
\centering
\caption{Apparent over-diffusion coefficient $\mathcal{R}_{\textrm{app}}$ for a water-filled aluminum foam at various macroscopic sample thicknesses.}
\renewcommand{\arraystretch}{1.5}
\begin{tabular}{l c}
$L_{\textrm{sample}}$ (mm) & $\mathcal{R}_{\textrm{app}}$ \\ \hline
2 & 15.5 \\
5 & 16.0 \\
20 & 16.1 \\
100 & 16.1 \\ \hline
\end{tabular} \label{tab:water_foam}
\end{table}

\subsubsection*{Case 3: Air-filled carbon foam}
Carbon foams exhibit high solid-phase thermal conductivity but relatively low density compared to metals (see the sample's details in \cite{FehEtal24}). Assuming a volume fraction of $\varepsilon_s = 0.20$, and a graphitic carbon structure, resulting in $C_s = 308 \, \mathrm{kJ/(m^3 \ K)}$, $k_s = 10 \, \mathrm{W/(m \ K)}$, and $C_f = 960 \, \mathrm{J/(m^3 \ K)}$, $k_f = 0.02 \, \mathrm{W/(m \ K)}$. Therefore, the static thermal conductivity is $k_{\textrm{eff}} = 10.02 \, \mathrm{W/(m \ K)}$, and the over-diffusion coefficient is $\mathcal{R} = 1.64$, which aligns with typical values from experimental observations.

In order to evaluate the finite boundary effects, we assume the same pore geometry as in the first case, retaining the volumetric heat transfer coefficient for air at $H = 546 \, \mathrm{kW/(m^3 \ K)}$. Because the effective solid and fluid conductivities are higher than in the air-filled aluminum foam, their product yields a notably higher spatial non-locality coefficient, $\mathcal{D} \approx 3.66 \cdot 10^{-7} \, \mathrm{W \ m / K}$. As a result, the higher-order spatial correction $\frac{\mathcal{D} \pi^2}{k_{\textrm{eff}} L_{\textrm{sample}}^2}$ plays a stronger role here. Table \ref{tab:carbon_foam} presents the apparent over-diffusion coefficient $\mathcal{R}_{\textrm{app}}(L_{\textrm{sample}})$, revealing a substantial drop in the over-diffusion behavior for samples thinner than $5 \, \mathrm{mm}$.

\begin{table}[h]
\centering
\caption{Apparent over-diffusion coefficient $\mathcal{R}_{\textrm{app}}$ for an air-filled carbon foam at various macroscopic sample thicknesses.}
\renewcommand{\arraystretch}{1.5}
\begin{tabular}{l c}
$L_{\textrm{sample}}$ (mm) & $\mathcal{R}_{\textrm{app}}$ \\ \hline
2 & 1.51 \\
5 & 1.62 \\
20 & 1.64 \\
100 & 1.64 \\ \hline
\end{tabular} \label{tab:carbon_foam}
\end{table}

\subsubsection*{Case 4: Water-saturated limestone}
Although rock thermal properties can significantly vary, we use $C_s = 1.8 \cdot 10^6 \, \mathrm{J/(m^3 \ K)}$, $k_s = 2 \, \mathrm{W/(m \ K)}$, $C_f = 836 \, \mathrm{kJ/(m^3 \ K)}$, $k_f = 0.12 \, \mathrm{W/(m \ K)}$, and assuming $\varepsilon_f = 0.20$ to test the developed theoretical background. The static conductivity is $k_{\textrm{eff}} = 2.12 \, \mathrm{W/(m \ K)}$. Although water has high heat capacity, the asymmetry is not that significant compared to Case 2, and thus we obtain $\mathcal{R} \approx 1.56$, which is in complete agreement with \cite{FehEtal21}.

In order to investigate the macroscopic size-dependence, we must establish a realistic volumetric heat transfer coefficient for the porous rock. Referencing the typical macroscopic relaxation time of $\tau_q \approx 0.5 \, \mathrm{s}$ observed in transient heat pulse experiments \cite{FehEtal21, FehKov24}, we can deduce an effective $H \approx 1.14 \cdot 10^6 \, \mathrm{W/(m^3 \ K)}$. This yields a spatial non-locality coefficient of $\mathcal{D} = k_s k_f / H \approx 2.1 \cdot 10^{-7} \, \mathrm{W \ m / K}$. For this natural heterogeneous material, the boundary effects severely dampen the over-diffusive response in very thin samples, as shown in Table \ref{tab:limestone}. At $L_{\textrm{sample}} = 2 \, \mathrm{mm}$, the apparent over-diffusion coefficient drops substantially to roughly $1.25$, demonstrating why non-Fourier behavior might be heavily suppressed or appear highly scale-dependent in micro-scale geological samples. However, in a heat pulse experiment, a small amount of heat is typically used to keep the heat conduction process linear; therefore, smaller samples are required. As we can observe, smaller samples can notably reduce the over-diffusion effects.   

\begin{table}[h]
\centering
\caption{Apparent over-diffusion coefficient $\mathcal{R}_{\textrm{app}}$ for a water-saturated limestone at various macroscopic sample thicknesses.}
\renewcommand{\arraystretch}{1.5}
\begin{tabular}{l c}
$L_{\textrm{sample}}$ (mm) & $\mathcal{R}_{\textrm{app}}$ \\ \hline
2 & 1.25 \\
5 & 1.50 \\
20 & 1.56 \\
100 & 1.56 \\ \hline
\end{tabular} \label{tab:limestone}
\end{table}

\subsubsection*{Case 5: Metal-organic frameworks (MOFs) with reduced graphene oxide (RGO)}

Metal-organic frameworks (MOFs) are characterized by nanometer-scale pores and poor thermal conductivity \cite{James03, DomEtal21, LiuEtal16}, which limit their industrial application in gas storage, where heat conduction can be important \cite{GalEtal24}. To improve their thermal properties, highly conductive secondary components, such as reduced graphene oxide (RGO), are mechanically mixed and compressed into the MOF matrix \cite{GalEtal24}. Recent heat pulse experiments conducted on compressed HKUST-1-RGO pellets (thicknesses between 0.68 and 1.15 mm) revealed interesting insights regarding the origin of non-Fourier behavior in these composites \cite{GalEtal24}. 

Notably, pure MOF pellets (0 mass\% RGO) exhibited classical diffusive behavior; the non-Fourier over-diffusive behavior, necessitating a two-timescale generalized heat equation, only emerged in the composites containing RGO \cite{GalEtal24}. Consequently, the effective two-phase continuum causing the deviation from Fourier's law is not the solid-air interaction, but rather the thermal interaction between the MOF matrix (phase $\alpha$) and the RGO additive (phase $\beta$).

We assign an effective macroscopic volumetric heat capacity for the MOF matrix as $C_\alpha \approx 2.5 \cdot 10^6 \, \mathrm{J/(m^3 \ K)}$. Because the MOF matrix retains its nanometer-scale porosity filled with trapped air, its effective phase conductivity is severely restricted to $k_\alpha \approx 0.07 \, \mathrm{W/(m \ K)}$ \cite{GalEtal24}. For the dispersed RGO phase, we approximate $C_\beta \approx 1.0 \cdot 10^6 \, \mathrm{J/(m^3 \ K)}$. Although individual RGO particles possess exceptionally high intrinsic thermal conductivity, their effective macroscopic contribution to the phase conductivity ($k_\beta$) remains restricted. This is because the RGO forms a multilayer structure rather than a continuous high-conducting material due to the presence of significant thermal resistance within the pellet \cite{GalEtal24}. Therefore, we estimate an effective value of $k_\beta \approx 0.35 \, \mathrm{W/(m \ K)}$, introducing a higher, yet notably bounded, conductivity contribution into the mixture. This yields an effective heat capacity $C_{\textrm{eff}} = 3.5 \cdot 10^6 \, \mathrm{J/(m^3 \ K)}$ and a static thermal conductivity $k_{\textrm{eff}} = 0.42 \, \mathrm{W/(m \ K)}$. The estimated thermal diffusivity is approximately $a_{\textrm{eff}} \approx 1.20 \cdot 10^{-7} \, \mathrm{m^2/s}$, which aligns with the magnitude of the experimental Fourier evaluations \cite{GalEtal24}. Despite this agreement, we note again that the studied MOF structure is highly complex; no separate (component-wise) measurements were performed; thus, these are mere estimates.

In the heat pulse experiments, the observed heat flux relaxation time for the RGO composites was consistently found to be $\tau_q \approx 0.45 \, \mathrm{s}$ \cite{GalEtal24}. By recalling the definition of the relaxation time $\tau_q = C_\alpha C_\beta / (H C_{\textrm{eff}})$, we can determine the volumetric heat transfer coefficient between the MOF matrix and the RGO, yielding $H \approx 1.59 \cdot 10^6 \, \mathrm{W/(m^3 \ K)}$. Because the samples are compressed pellets, this interfacial thermal coupling can be physically realistic.

Evaluating the intrinsic, scale-independent over-diffusion coefficient for this structural configuration yields a value of $\mathcal{R} \approx 3.15$. However, it is essential to highlight the size-dependence observed in the sub-RVE experimental samples. For a sample thickness of $L_{\textrm{sample}} \approx 0.7 \, \mathrm{mm}$, the spatial non-locality coefficient is $\mathcal{D} = k_\alpha k_\beta / H \approx 1.54 \cdot 10^{-8} \, \mathrm{W \ m / K}$, and by applying Eq.~\eqref{eq:R_ratio_exact}, the apparent over-diffusion coefficient drops substantially from its intrinsic value, yielding $\mathcal{R}_{\textrm{app}} \approx 1.81$. Table \ref{tab:mof} illustrates this significant scale dependence, confirming that small macroscopic samples inherently suppress the full realization of the over-diffusive regime.

\begin{table}[h]
\centering
\caption{Apparent over-diffusion coefficient $\mathcal{R}_{\textrm{app}}$ for a compressed metal-organic framework/RGO pellet at various macroscopic sample thicknesses.}
\renewcommand{\arraystretch}{1.5}
\begin{tabular}{l c}
$L_{\textrm{sample}}$ (mm) & $\mathcal{R}_{\textrm{app}}$ \\ \hline
0.7 & 1.81 \\
1.0 & 2.31 \\
2.0 & 2.89 \\
10.0 & 3.14 \\
100.0 & 3.15 \\ \hline
\end{tabular} \label{tab:mof}
\end{table}

\section{Extensions of the LTNE Framework}

\subsection{Application to biological heat transfer}

Classical bio-heat transfer is usually modeled using the Pennes heat equation \cite{Penn48}, which treats living tissue as a homogenized single-temperature continuum and models blood perfusion as an isotropic heat source. In other words, the Pennes heat equation uses Fourier's law and introduces various heat sources in order to include biological effects (e.g., blood perfusion in this case) mathematically. This approach assumes local thermal equilibrium between the capillary network and the surrounding cellular matrix. However, during rapid, high-energy medical interventions \cite{Kutz09b, BecKuz14b, Hoosetal15}, the localized heating rate can be immense. Under significant transients, the blood network and the solid tissue can thermally decouple, behaving analogously to a two-temperature microscopic system. 

We model the biological tissue as a two-phase porous structure:
\begin{itemize}
    \item Solid phase ($\alpha$): the extravascular tissue matrix, with volume fraction $\varepsilon_\alpha$.
    \item Fluid phase ($\beta$): the intravascular blood flowing through the capillary network, with volume fraction $\varepsilon_\beta$ (where $\varepsilon_\alpha + \varepsilon_\beta = 1$).
\end{itemize}
Applying the spatial averaging theorem to the phase energy balances, we recover the coupled macroscopic equations:
\begin{align}
    C_\alpha \frac{\partial T_\alpha}{\partial t} &= \nabla \cdot (k_\alpha \nabla T_\alpha) - H(T_\alpha - T_\beta) + q_{met}, \label{eq:bio_tissue} \\
    C_\beta \frac{\partial T_\beta}{\partial t} + C_\beta \mathbf{u}_\beta \cdot \nabla T_\beta &= \nabla \cdot (k_\beta \nabla T_\beta) + H(T_\alpha - T_\beta) - W_\beta (T_\beta - T_{art}), \label{eq:bio_blood}
\end{align}
where $q_{met}$ is metabolic heat generation, $\mathbf{u}_\beta$ is the macroscopic directional blood velocity, $W_\beta = \omega_\beta (\rho c)_\beta$ in which $\omega_\beta$ is the scalar blood perfusion rate, and $T_{art}$ is the arterial temperature. In subsequent calculations, we neglect macroscopic advection ($\mathbf{u}_\beta = \mathbf{0}$) for simplicity, but we note that this can be an oversimplification in numerous situations.  

In order to recover the modified over-diffusion coefficient, we need to eliminate the individual phase temperatures and obtain the continuum behavior for the macroscopic mixture temperature $T = (C_\alpha T_\alpha + C_\beta T_\beta)/C_{\textrm{eff}}$. After some algebra, we obtain the generalized macroscopic differential equation,
\begin{align}
    \frac{C_\alpha C_\beta}{H} \frac{\partial^2 T}{\partial t^2} + C_{\textrm{app}}^* \frac{\partial T}{\partial t} =k_{\textrm{app}}^* \Delta T + \frac{C_\alpha k_\beta + C_\beta k_\alpha}{H} \frac{\partial \Delta T}{\partial t} - \frac{k_\alpha k_\beta}{H} \Delta \Delta T 
     - W_\beta (T - T_{art}) + S_{\textrm{eff}}, \label{eq:exact_bioheat}
\end{align}
where the apparent macroscopic volumetric capacity $C_{\textrm{app}}^*$ and static conductivity $k_{\textrm{app}}^*$ are defined as:
\begin{align}
    C_{\textrm{app}}^* = C_{\textrm{eff}} + \frac{W_\beta}{H} C_\alpha, \quad 
    k_{\textrm{app}}^* = k_{\textrm{eff}} + \frac{W_\beta}{H} k_\alpha, \label{eq:k_app_perf}
\end{align}
and the generalized effective source term $S_{\textrm{eff}}$, which accounts for the temporal and spatial variations of the metabolic heat generation and the arterial temperature, reads
\begin{align}
    S_{\textrm{eff}} =   \frac{C_{\textrm{app}}^*}{C_{\textrm{eff}}} q_{met} + \frac{C_\alpha C_\beta}{H C_{\textrm{eff}}} \left( \frac{\partial q_{met}}{\partial t} + W_\beta \frac{\partial T_{art}}{\partial t} \right) 
     - \frac{1}{H C_{\textrm{eff}}} \left( C_\alpha k_\beta \Delta q_{met} + C_\beta k_\alpha W_\beta \Delta T_{art} \right). \label{eq:bio_source}
\end{align}
For constant $q_{met}$, Eq.~\eqref{eq:bio_source} greatly simplifies. We can identify the modified macroscopic transport parameters based on Equation \eqref{eq:exact_bioheat}, $\tau_q^* = C_\alpha C_\beta / (H C_{\textrm{app}}^*)$ and $\tau{_T^{}}^* = (C_\alpha k_\beta + C_\beta k_\alpha) / (H k_{\textrm{app}}^*)$. We can evaluate the ratio ($\tau_{T^{}}^* / \tau_q^*$) to express the difference between the static and dynamic thermal behavior, hence we obtain a perfusion-augmented over-diffusion coefficient in the form,
\begin{equation}
    \mathcal R_{bio}(W_\beta) = \left( \frac{C_\alpha k_\beta + C_\beta k_\alpha}{C_\alpha C_\beta} \right) \left( \frac{H C_{\textrm{eff}} + W_\beta C_\alpha}{H k_{\textrm{eff}} + W_\beta k_\alpha} \right). \label{eq:M_bio_perfusion}
\end{equation}

\subsubsection{Clinical implications of over-diffusive tissue}

The mathematical proof that living tissue is fundamentally over-diffusive, meaning $\mathcal{R}_{bio} > 1$, carries critical consequences for surgical planning. Classical Pennes and Fourier models can systematically fail to predict the early transient thermal penetration during high-energy procedures.

As a first example, we consider cancellous bone and spinal radiofrequency ablation. Radiofrequency ablation is routinely used to destroy metastatic tumors within vertebrae \cite{DupuyEtal00, MasEtal04}. The clinical challenge is ensuring the destruction zone concentrates on the tumor without the thermal front affecting the posterior vertebral wall, which causes irreversible necrosis to the spinal cord. Cancellous bone consists of a solid matrix of calcified trabeculae (phase $\alpha$) with a volume fraction of $\varepsilon_\alpha = 0.20$, and a fluid phase of highly vascularized bone marrow (phase $\beta$) with $\varepsilon_\beta = 0.80$. The solid trabeculae possess relatively high thermal conductivity, leading to an effective phase conductivity of $k_\alpha = 0.13 \, \mathrm{W/(m \ K)}$, with a low effective volumetric capacity of $C_\alpha = 5.0 \cdot 10^5 \, \mathrm{J/(m^3 \ K)}$. The vascular marrow has an effective phase conductivity, $k_\beta = 0.21 \, \mathrm{W/(m \ K)}$, but a much larger effective heat capacity of $C_\beta = 3.0 \cdot 10^6 \, \mathrm{J/(m^3 \ K)}$. The static thermal conductivity of the bone is $k_{\textrm{eff}} = 0.34 \, \mathrm{W/(m \ K)}$. Assuming localized behavior without strong perfusion effects ($W_\beta \approx 0$), evaluating the biological over-diffusion coefficient results in
\begin{equation}
    \mathcal{R}_{spine} = 1 + \frac{k_\alpha C_\beta^2 + k_\beta C_\alpha^2}{C_\alpha C_\beta k_{\textrm{eff}}} \approx 3.40.
\end{equation}
Because $\mathcal{R}_{spine} \approx 3.40$, the initial thermal response propagates through the calcified trabeculae more than three times faster than classical Fourier theory predicts. The affected area can be even larger if $W_\beta \neq 0$, increasing $\mathcal{R}_{spine}$. On the other hand, it can increase the effective heat transfer coefficient $H$, thereby leading to non-trivial outcomes.

As a second example, we investigate the dermis and cutaneous laser surgery. In dermatology, ultra-short pulsed lasers aim for selective photothermolysis, destroying a target before heat diffuses into surrounding healthy skin. The dermis features a relatively large thermal inertia solid matrix of collagen and elastin (phase $\alpha$) with $\varepsilon_\alpha = 0.95$, including a micro-vascular capillary plexus (phase $\beta$) with $\varepsilon_\beta = 0.05$. The solid dermis has parameters $C_\alpha = 3.42 \cdot 10^6 \, \mathrm{J/(m^3 \ K)}$ and $k_\alpha = 0.35 \, \mathrm{W/(m \ K)}$. The micro-vascular plexus has $C_\beta = 2.1 \cdot 10^5 \, \mathrm{J/(m^3 \ K)}$ and $k_\beta = 0.10 \, \mathrm{W/(m \ K)}$. The static effective conductivity is $k_{\textrm{eff}} = 0.45 \, \mathrm{W/(m \ K)}$. Evaluating the over-diffusion coefficient,
\begin{equation}
    \mathcal{R}_{skin} = 1 + \frac{k_\alpha C_\beta^2 + k_\beta C_\alpha^2}{C_\alpha C_\beta k_{\textrm{eff}}} \approx 4.67,
\end{equation}
which shows notable dynamic behavior. If the pulse duration approaches the relaxation time, the significant dynamic conductivity dominates, leading to a rapid non-Fourier thermal transient that affects the dermal matrix, potentially causing collateral thermal damage, and demonstrating that the generalized Jeffreys model can be relevant for optimizing laser pulse durations and other biological applications.

By defining $\mathcal{R}_{bio}$ and the augmented properties analytically based on the local vascular volume fraction taken from medical perfusion scans, one could estimate a more precise non-Fourier thermal behavior and avoid unnecessary tissue necrosis.

\subsection{The role of thermal radiation in over-diffusion}

In highly porous, high-temperature heterogeneous media, such as open-cell carbon and metal foams, heat transport is governed by the simultaneous presence of solid and fluid conduction, as well as thermal radiation, across pore cavities. So far, we have neglected radiation in transient analyses, but its inclusion can be essential for accurately capturing the over-diffusive regime where $\mathcal{R} > 1$.

At the microscale, thermal radiation acts as an ultrafast parallel heat transfer channel, thermally connecting opposite pore walls. To incorporate this into the macroscopic continuum framework without solving complex integro-differential radiative transfer equations, radiation across a pore of characteristic size $l_{pore}$ is linearized into an equivalent radiative heat transfer coefficient,
\begin{equation}
    h_{rad} = 4 \sigma \epsilon_{\textrm{eff}} \bar{T}^3,
\end{equation}
where $\sigma = 5.67 \cdot 10^{-8} \, \mathrm{W/(m^2 \ K^4)}$ is the Stefan-Boltzmann constant, $\epsilon_{\textrm{eff}}$ is the effective cavity wall emissivity, and $\bar{T}$ is the mean absolute temperature. Radiative exchange can notably increase the total volumetric interfacial heat transfer coefficient that governs the heat flux relaxation time. The total volumetric exchange is $H = H_{cond} + H_{rad}$, where $H_{rad} \approx h_{rad} / l_{pore}$ represents the volumetric radiative exchange rate. Because radiation introduces an immediate parallel mechanism for energy transfer across pores, the augmented $H$ directly decreases the relaxation time,
\begin{equation}
    \tau_q = \frac{C_\alpha C_\beta}{(H_{cond} + H_{rad}) C_{\textrm{eff}}}.
\end{equation}
We note that $H$ does not directly modify the over-diffusion coefficient $\mathcal{R}$, thus the radiation effects do not affect the ratio of the static and dynamic behavior. However, the increased volumetric exchange plays a profound role in the apparent, scale-dependent over-diffusion coefficient $\mathcal{R}_{\textrm{app}}$ observed in finite samples. By increasing $H$, radiation can notably reduce the spatial non-locality coefficient, defined as $\mathcal{D} = k_\alpha k_\beta / H$. This decrease suppresses the finite-boundary correction term, thereby driving the apparent over-diffusion coefficient $\mathcal{R}_{\textrm{app}}$ much closer to its theoretical maximum.

Furthermore, thermal radiation also acts as a spatial parallel transport mechanism, which directly augments the effective macroscopic static conductivity of the fluid pore phase $k_\beta$. Because the solid matrix possesses a much larger thermal capacity than the gas-filled pores ($C_\alpha \gg C_\beta$), any radiative augmentation to $k_\beta$ significantly influences the $k_\beta C_\alpha^2$ term in the numerator of the intrinsic over-diffusion coefficient. Consequently, both the intrinsic theoretical limit $\mathcal{R}$ and the apparent $\mathcal{R}_{\textrm{app}}$ can show a pronounced relative increase. This radiation-mechanism enhancement can explain why highly porous carbon foams exhibit sharp initial transient overshoots that conduction-only models fail to resolve \cite{FehEtal24}.

\section{Discussion and Summary}

The rigorous upscaling of the present LTNE framework bridges the gap between microstructural geometry and macroscopic phenomenological heat conduction. By applying spatial volume averaging to a heterogeneous two-phase continuum, we have mathematically proven that non-Fourier thermal behavior is not only a phenomenological assumption but a direct consequence of thermal asymmetries between the constituents. The derivation yields a generalized Jeffreys-type heat equation. Consequently, any heterogeneous material modeled via this framework inherently falls into the over-diffusive regime ($\mathcal{R} > 1$). Furthermore, the model satisfies the Clausius-Duhem inequality, thereby providing an upper bound on the emergence of the non-local spatial term and ensuring positive entropy production. 

Furthermore, the present derivation has a non-trivial consequence. Most experiments on heterogeneous materials are evaluated using the continuum version of the Guyer--Krumhansl heat equation \cite{Botetal16, FehEtal21, FehEtal24b}. Such a model consists of the same mathematical operators; however, it is derived within the internal variable framework \cite{VanFul12} and is free from the constraints of phonon hydrodynamics. However, we have proved that under particular conditions, the Guyer--Krumhansl and Jeffreys heat equations can be equivalent in the sense that both equations model the same temperature history and distribution \cite{FehKov24}. These conditions restrict the volumetric heat sources; the heat flux must be curl-free, and all transport parameters must be constant, besides ensuring the same homogeneous equilibrium for the initial state and applying the same boundary conditions. Consequently, in the light of the present derivation, only these particular conditions enabled the use of the Guyer--Krumhansl model in the earlier experiments. 

We have demonstrated that higher-order spatial non-localities are sensitive to macroscopic boundary constraints in finite-thickness samples, suppressing the apparent over-diffusion in sub-RVE flash experiments. By comparing the structural thermal properties with numerous experimental results, the model provides a robust alternative to two-temperature formulations and explains why scale-dependent thermal behavior is frequently observed in thin porous media and composites.

Looking forward, the clinical implications of this over-diffusive heat conduction framework are profound for advanced biological and medical applications. Classical bio-heat models, such as the Pennes equation, assume instantaneous thermal equilibrium, which can fail to accurately predict transient thermal penetration during high-energy, short-duration interventions. In procedures such as radiofrequency spinal ablation, the non-Fourier thermal transients can propagate significantly faster than classical diffusion predicts. Implementing the perfusion-augmented structural multiplier ($\mathcal{R}_{bio}$) in surgical models may enable more accurate mapping of these thermal transients, thereby minimizing collateral damage and optimizing targeted thermal therapies.

\section{Acknowledgement}
The research reported in this paper is part of project no. TKP-6-6/PALY-2021, implemented with the support provided by the Ministry of Culture and Innovation of Hungary from the National Research, Development and Innovation Fund, financed under the TKP2021-NVA funding scheme. The research carried out at BME has been also supported by the grant National Research, Development and Innovation Office-NKFIH STARTING24 149487 and by the Sustainable Development and Technologies National Programme of the Hungarian Academy of Sciences (FFT NP FTA). This paper was supported by the János Bolyai Research Scholarship of the Hungarian Academy of Sciences, Hungary.

\bibliographystyle{unsrt}
\bibliography{bibliography}

\end{document}